\documentclass[12pt]{article}
\usepackage{amsmath}
\usepackage{graphicx}
\usepackage{comment}
\usepackage{natbib}
\usepackage{lmodern}
\usepackage{url}
\usepackage{wrapfig}
\usepackage{booktabs}
\usepackage{appendix}
\usepackage{amsmath,bm,amssymb,amsthm,setspace,paralist}
\usepackage[colorlinks,citecolor=blue,urlcolor=blue]{hyperref}
\usepackage{epstopdf}
\usepackage[normalem]{ulem}
\usepackage{multirow}
\usepackage{float}
\usepackage[inline]{enumitem}
\usepackage{bbm}
\usepackage{subcaption}
\usepackage{rotating}
\usepackage{standalone}
\usepackage{algorithm,algpseudocode}
\theoremstyle{plain}
\newtheorem{theorem}{Theorem}[section]
\newtheorem{lemma}[theorem]{Lemma}

\newtheorem{corollary}[theorem]{Corollary}
\usepackage{rotating}
\usepackage{tikz}
\usetikzlibrary{shapes,bayesnet}
\theoremstyle{plain}

\def\m{\mathcal}

\newcommand{\+}[1]{\ensuremath{\boldsymbol {#1}}}

\newcommand{\be}{\begin{equs}}
  \newcommand{\ee}{\end{equs}}
\newcommand{\bpm}{\begin{pmatrix}}
  \newcommand{\epm}{\end{pmatrix}}
\DeclareMathOperator{\E}{\mathbf E}

\newcommand{\ind}{\mathbbm 1}
\newcommand{\indm}{\mathbbm I}

\DeclareMathOperator*{\argmin}{argmin}

\usepackage[inline]{enumitem}

\newtheorem{assumption}{Assumption}
\usepackage{float}

\usepackage{accents}
\usepackage{bibunits}
\usepackage{tikz}
\usepackage{animate}
\usetikzlibrary{shapes,bayesnet}
\usepackage{bibunits}
\defaultbibliographystyle{apalike}
\defaultbibliography{main.bib}
\begin{document}

\def\spacingset#1{\renewcommand{\baselinestretch}%
{#1}\small\normalsize} \spacingset{1}

%%%%%%%%%%%%%%%%%%%%%%%%%%%%%%%%%%%%%%%%%%%%%%%%%%%%%%%%%%%%%%%%%%%%%%%%%%%%%%

  \title{\bf Robust High-Dimensional Covariate-Assisted Network Modeling}
  \author{Peng Zhao\\
    Department of Applied Economics and Statistics, University of Delaware \\
    Yabo Niu\\ Department of Mathematics, University of Houston}
    \date{}
  \maketitle

\bigskip

  \graphicspath{{figures/}}
  
\maketitle
  \bigskip

\begin{abstract}
Modern network data analysis often involves analyzing network structures alongside covariate features to gain deeper insights into underlying patterns. However, traditional covariate-assisted statistical network models may not adequately handle cases involving high-dimensional covariates, where some covariates could be uninformative or misleading, or the possible mismatch between network and covariate information. To address this issue, we introduce a novel robust high-dimensional covariate-assisted latent space model. This framework links latent vectors representing network structures with simultaneously sparse and low-rank transformations of the high-dimensional covariates, capturing the mutual dependence between network structures and covariates. To robustly integrate this dependence, we use a shrinkage prior on the discrepancy between latent network vectors and low-rank covariate approximation vectors, allowing for potential mismatches between network and covariate information. For scalable inference, we develop two variational inference algorithms, enabling efficient analysis of large-scale sparse networks. We establish the posterior concentration rate within a suitable parameter space and demonstrate how the proposed model facilitates adaptive information aggregation between networks and high-dimensional covariates. Extensive simulation studies and real-world data analyses confirm the effectiveness of our approach.
\end{abstract}

  \noindent%
  \vfill
\newpage
\spacingset{1.8}

%    \begin{bibunit}
\section{Introduction}

Network data refers to the relationships or interactions among individuals, usually represented by nodes and edges. A node represents an individual, and an edge represents the interaction between the two nodes it connects. The statistical analysis of network-valued data is gaining popularity in modern scientific research. It has a broad range of real-world applications, including authorship identifications in word co-occurrence networks \citep{akimushkin2017text}, social hierarchy detection in email communication networks \citep{rowe2007automated}, and the study of disease spreading processes in physical proximity networks \citep{stopczynski2018physical}. For reviews, please refer to \cite{goldenberg2010survey}, \cite{snijders2011statistical}, and \cite{newman2018networks}. In numerous scenarios, alongside network data, additional high-dimensional covariates for nodes are gathered. For example, in healthcare networks, where patients are represented by nodes and their shared medical histories are represented by edges, patients are associated with high-dimensional information such as demographics, biometrics and clinical assessments. Research has consistently shown that these node attributes provide complementary insights to network links and play a pivotal role in uncovering latent structures of networks \citep{newman2016structure,niu2023covariate}. Consequently, it is important to model the network jointly with covariate information, which benefits downstream network analysis tasks such as community detection \citep{fortunato2016community} and link predictions \citep{martinez2016survey}.

Several works in the literature have modeled networks with the help of covariate information. Based on spectral clustering, \cite{binkiewicz2017covariate} and \cite{hu2024network} proposed covariate-assisted spectral clustering approaches for network data. In general, they considered applying spectral clustering on the following weighted matrix between the network and covariates:
\begin{equation}\label{eq:literature}
\widetilde{\+{Y}} =  \+{Y} +\alpha K(\+{Z},\+{Z}),
\end{equation}
where $\+{Y} \in \mathbb{R}^{n \times n}$ with 0/1 entries is the original or some variant form (e.g., \+{Y}\+{Y'}) of the adjacency matrix or Laplacian matrix, $\alpha \in \mathbb{R}$ is a tuning parameter to balance between the network and covariate information, and $K(\cdot,\cdot)$ is a kernel function operating on the covariates $\+Z \in \mathbb{R}^{n \times p}$, e.g.,  $K(\+{Z},\+{Z})=\+Z\+Z'$. Although they have improved upon the spectral method that only relies on network information, their methods utilize all covariates without selection and only a \textit{uniform} parameter across all nodes is used to determine whether to receive information from the covariates.  For more details, refer to Section 2.2 of \cite{binkiewicz2017covariate} and Section 2.4 of \cite{hu2024network}. In addition to spectral methods, various non-spectral approaches have been developed for covariate-assisted network modeling. \cite{yan2019statistical} utilizes all covariates to construct a pairwise matrix, but assumes fixed covariate dimensionality, limiting its applicability in high-dimensional settings. \cite{zhang2016community} employs a single parameter to balance the contributions of network structure and covariates (see their Section A.2). \cite{yan2021covariate} applies a semi-definite programming approach to solve a $K$-means type problem for the weighted matrix~\eqref{eq:literature}. \cite{zhang2022joint} constructs latent space models that have the same latent vectors for networks and covariates. However, this method employs only a single tuning parameter to balance the likelihood contributions from network and covariates and does not address the problem of sparse high-dimensional covariate effects.

\begin{figure}
  \centering
  \scalebox{0.75}{
  \begin{tikzpicture}[thick]

  \shade[black] (0,0) -- (5,0) -- (7,2) -- (2,2) -- cycle;
  \node[] at(7, 1) {$z=0$};
  \shade[black] (0,2) -- (5,2) -- (7,4) -- (2,4) -- cycle;
  \node[] at(7, 3.25) {$z=1$};
  \shade[black] (0,-3.5) -- (5,-3.5) -- (7,-1.5) -- (2,-1.5) -- cycle;
  \node[] at(8, -2.5) {network space};
  \node[] at(8, 2.25) {covariate space};

  \node[draw, shape=circle, fill=black, inner sep=1pt] at (1,-3) (x1) {};
  \node[draw, shape=circle, fill=gray, inner sep=1pt] at (1,0.5) (x10) {};
  \edge[-, dashed] {x1} {x10};

  \node[draw, shape=circle, fill=black, inner sep=1pt] at (1.5,-2.5) (x2) {};
  \node[draw, shape=circle, fill=gray, inner sep=1pt] at (1.5,1) (x20) {};
  \edge[-, dashed] {x2} {x20};

  \node[draw, shape=circle, fill=black, inner sep=1pt] at (3,-2.5) (x3) {};
  \node[draw, shape=circle, fill=gray, inner sep=1pt] at (3,1) (x30) {};
  \edge[-, dashed] {x3} {x30};

  \node[draw, shape=circle, fill=black, inner sep=1pt] at (2,-3) (x4) {};
  \node[draw, shape=circle, fill=gray, inner sep=1pt] at (2,0.5) (x40) {};
  \edge[-, dashed] {x4} {x40};

  \node[draw, shape=circle, fill=black, inner sep=1pt] at (2.75,-2) (x5) {};
  \node[draw, shape=circle, fill=green, inner sep=1pt] at (2.75,3.5) (x51) {};
  \edge[-, dashed] {x5} {x51};

  \edge[color = blue, -] {x1} {x2};
  \edge[color = blue, -] {x1} {x3};
  \edge[color = blue, -] {x1} {x4};
  \edge[color = blue, -] {x1} {x5};
  \edge[color = blue, -] {x2} {x3};
  \edge[color = blue, -] {x2} {x4};
  \edge[color = blue, -] {x2} {x5};
  \edge[color = blue, -] {x3} {x4};
  \edge[color = blue, -] {x3} {x5};
  \edge[color = blue, -] {x4} {x5};

  \node[draw, shape=circle, fill=black, inner sep=1pt] at (6,-2) (y1) {};
  \node[draw, shape=circle, fill=black, inner sep=1pt] at (6,3.5) (y11) {};
  \edge[-, dashed] {y1} {y11};

  \node[draw, shape=circle, fill=black, inner sep=1pt] at (5,-3) (y2) {};
  \node[draw, shape=circle, fill=black, inner sep=1pt] at (5,2.5) (y21) {};
  \edge[-, dashed] {y2} {y21};

  \node[draw, shape=circle, fill=black, inner sep=1pt] at (4,-2.5) (y3) {};
  \node[draw, shape=circle, fill=black, inner sep=1pt] at (4,3) (y31) {};
  \edge[-, dashed] {y3} {y31};

  \node[draw, shape=circle, fill=black, inner sep=1pt] at (4.5,-2) (y4) {};
  \node[draw, shape=circle, fill=yellow, inner sep=1pt] at (4.5,1.5) (y40) {};
  \edge[-, dashed] {y4} {y40};

  \edge[color = red, -] {y1} {y2};
  \edge[color = red, -] {y1} {y3};
  \edge[color = red, -] {y1} {y4};
  \edge[color = red, -] {y2} {y3};
  \edge[color = red, -] {y2} {y4};
  \edge[color = red, -] {y3} {y4};
  \end{tikzpicture}}
  \caption{A toy example showing the mismatch between binary covariate values and cluster allocation of nodes in the network. The dashed lines indicate the covariate values corresponding to network nodes. Red and blue edges denote the membership of network clusters.} \label{fig:concept}
\end{figure}
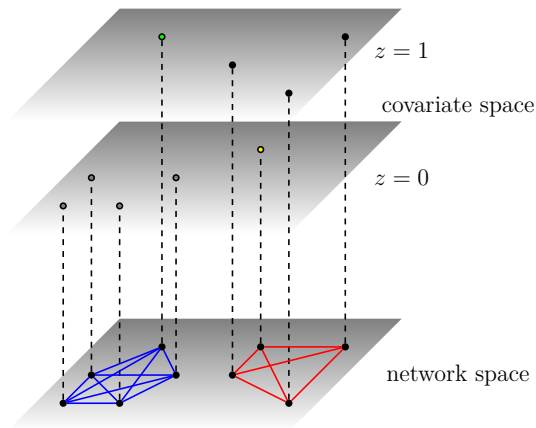

When analyzing real datasets, two critical challenges arise in covariate-assisted network modeling: 1) High-dimensional covariate effects often exhibit sparsity, where covariate dimensionality approaches sample size and only a subset of covariates meaningfully contributes to network modeling; 
2) There could be a mismatch between node-level information derived from networks and information extracted from covariates. This mismatch commonly occurs because networks and covariate observations frequently originate from \textit{different sources} or represent \textit{distinct aspects} of the subjects under study. For instance, in the LastFM Asia social network dataset \citep{rozemberczki2020characteristic}, users' friendship connection networks with latent communities based on countries do not consistently align with covariates indicating preferences for international artists. Similarly, in congressional voting networks, lawmakers frequently vote across party lines \citep{hager2000look}. Figure~\ref{fig:concept} illustrates this mismatching information with a toy example featuring a network and binary covariate. The binary covariate values do not perfectly correspond to the network's cluster structure: one node in the blue-edged network cluster possesses a covariate value (indicated by the green dot) that differs from other nodes in its cluster, while a node in the red-edged cluster displays a similar discrepancy (denoted by the yellow dot). Existing approaches that apply uniform modeling across all nodes and covariates fail to address this fundamental mismatch
%Such mismatch in the toy example happens often in real life. 

% \begin{figure}[H]
%  \centering
%     \includegraphics[width=0.5\linewidth]{concept_fig.jpeg}
%     \caption{ Visual representation showing the mismatch between the data at planes $Z=1$ (covariate) and $Z=0$ (node positions). The black lines show the corresponding of data between $Z=1$ and $Z=0$. The red and blue colors indicate the cluster membership of the data at the plane $Z=1$. The red and lightblue lines at $Z=0$ are hypothetical observed links. }\label{fig:concept}
% \end{figure} 

To address these two practical challenges, we propose a covariate-assisted latent space model for networks that implicitly selects relevant high-dimensional covariates. In contrast to existing approaches that model networks and covariates comprehensively, our method distinctively represents the effects of nodes and covariates as separate components within the latent space framework. Employing a Bayesian methodology, we implement individualized priors (global-local shrinkage, e.g., \cite{carvalho2010horseshoe}) on both covariates and network nodes. This approach parallels global-local shrinkage techniques in high-dimensional regression models, offering enhanced flexibility in determining which covariates significantly influence the network and which nodes should incorporate covariate information. Our methodological contributions can be divided into two key aspects utilizing shrinkage priors: 1) introduction of sparsity in covariate effects, enabling implicit selection of only the most relevant predictors; and 2) accommodation of potential mismatch between network structure and covariate information, facilitating robust information integration from covariates when appropriate.

From a methodological perspective, our proposed model offers two significant advantages. First, by assuming sparsity in the relationship between latent vectors and covariates, it achieves better network estimation compared to methods that use all covariates indiscriminately. This advantage is particularly valuable when only a subset of high-dimensional covariates contains useful information, which is a common scenario in practice. Second, the model adapts flexibly to diverse real-world situations: when network structure aligns with covariate data, it effectively combines both sources to learn latent vectors; conversely, when network information conflicts with covariates for certain nodes, it selectively prioritizes network data for those specific nodes. This adaptive capability extends to extreme cases where covariates provide no useful information across all nodes, equivalent to scenarios with zero non-zero coefficients for covariates. In such cases, the model naturally reverts to learning only from network structure. Consequently, the proposed approach achieves a balance between relying on the network structure for robustness and leveraging additional information to enhance network inference in a \textit{nodewise adaptive} manner. 

%Furthermore, despite considering all nodes and covariate variables individually, the incorporation of shrinkage priors within the Bayesian framework eliminates the need for tuning parameters (except for latent dimensionality) due to marginalization of high-level scale parameters. 

From a computational perspective, traditional sampling-based posterior inference becomes prohibitively inefficient when handling networks with a large number of nodes. To address this limitation, posterior approximations based on mean-field variational inference have been developed \citep{sewell2017latent,liu2022variational}, where the variational posteriors of all latent positions are assumed to be independent. This paper introduces two variational inference algorithms: 1) a coordinate ascent variational inference (\cite{bishop2006pattern}, CAVI) algorithm with computational complexity of $O(n^2+p^2)$ per iteration, using Rcpp implementation, offering convenient practical application for small-scale dense networks using Rcpp implementation. By leveraging tangent transformation \citep{jaakkola2000bayesian} of Bernoulli likelihood and Inverse-Gamma mixtures representation of the employed global-local prior \citep{neville2014mean}, we derive the CAVI algorithm in a conditional conjugate form. The detailed algorithm specification is provided in Section~\ref{sec:CAVI} of the supplement. 2) For large-scale sparse networks, we complement the CAVI approach with an alternative stochastic variational inference (SVI) algorithm implemented in PyTorch. This algorithm utilizes mini-batch sampling techniques to process only a subset of edges at each iteration, and an unbiased estimation is constructed to approximate the likelihood. The computational complexity of evaluating the likelihood function is significantly reduced to $O(|E|)$, where $|E|$ represents the number of positive edges in the network.  The complete specification of this algorithm, including mini-batch selection strategies and selection of variational distributions, is provided in data analysis Section~\ref{sec:real_data}.

From a theoretical perspective, statistical analysis of \textit{structured} latent space models, where latent vectors are constrained to specific parameter spaces, has garnered significant attention recently \citep{zhang2020flexible,zhang2022joint, zhao2022structured, zhao2022factorized, loyal2024fast}. These structured latent space models, which impose dependence structures among network latent positions, demonstrate improved theoretical posterior convergence rates compared to their non-structured counterparts.  In this paper, under the homogeneous sparse network setting, we analyze fractional posteriors through the lens of $\alpha$-R\'{e}nyi divergences. We derive the convergence rate of Bayes risk bound under the fractional posterior for recovering connection probabilities in an appropriate metric by establishing a novel concentration inequality for multivariate shrinkage priors. We demonstrate that this convergence rate improves upon that of latent space models without covariate information, potentially enabling superior misclustering rates or requiring less stringent separation conditions between communities for effective community detection.

%motivated by the recent development of Bayesian oracle inequalities for $\alpha$-R\'{e}nyi divergence risks \citep{bhattacharya2019bayesian}, 
The rest of this paper is organized as follows. The model structure, proposed priors and their properties are given in Section~\ref{sec:method}. Section~\ref{sec:theory} provides the theoretical results for contraction of the posterior and community detection.  Section~\ref{sec:computation} provides the proposed variational inference algorithm. Section~\ref {sec:data_analysis}  provides simulation studies and analysis of the real-world dataset
using the proposed method, respectively. We conclude this paper with discussions in Section~\ref{sec:discussion}.  All technical
proofs and detailed steps of proposed algorithms are provided in the Supplementary Materials.

\textbf{Notation.} For a vector $\+x$, we use $\|\+x\|_2$, $\|\+x\|_1$, $\|\+x\|_\infty$ to represent its $\ell_2$, $\ell_1$ and $\ell_\infty$ norms and $\+x'$ as its transpose.  For an $n \times m$ matrix $\+A=(\+a_1,...,\+a_m)$, let $\|\+A\|_F$ be its Frobenius norm and $\|\+A\|_{p,q}$ be its $L_{p,q}$ norm: $\|\+A\|_{p,q} =  (\sum_{j=1}^m (\sum_{i=1}^n |a_{i j}|^p )^{q/p} )^{1/q}$.  We use $\indm$  and $\+1$  to denote the identity matrix and vector with all ones. Suppose $P$ and $Q$ are probability measures on a common probability space with a dominating measure $\mu$,
and let $p = dP/d\mu, q = dQ/d\mu$. We use $D_{KL} \left\{ p \mid \mid q \right\} = \int p \log (p/q) d\mu $ to denote the KL divergence between the density $p$ and $q$. In addition, we use $D_{\alpha} \left\{ x\mid \mid x_0 \right\} = \log \int p_x^{\alpha} p_{x_0}^{1-\alpha} d\mu $ to denote the  R\'{e}nyi divergence of order $\alpha$ between the density $p_x$ and $p_{x_0}$.  Given sequences $a_n$ and $b_n$, we denote $a_n = O(b_n)$ or
	$a_n \lesssim b_n$ if there exists a constant $C>0$ such that $a_n \leq C b_n$ for all large enough $n$. Similarly, we
	define $a_n \gtrsim b_n$. In addition, let $a_n=o(b_n)$ and $a_n \ll b_n$ to denote $\lim_{n \rightarrow \infty} a_n/b_n = 0$. Let $P_X$ denote a probability distribution with parameter $X$, and $p_X$ denote the corresponding density function. Denote $\E_{x} $ as the expectation taken with respect to a variable $x$. Let $\mathcal{N}(\mu,\sigma)$ be the normal distribution with mean $\mu$ and variance $\sigma$ while $N(x;\mu,\sigma)$ be the corresponding density function at $x$. 
\section{Covariate-Assisted Latent Space Model}\label{sec:method}
\subsection{Proposed model}

We start by introducing the latent space model. Suppose we observe a network data $\+Y \in \mathbb{R}^{n \times n}$. Let true latent positions $\+X^*=[\+x_1,...,\+x_n]' \in \mathbb{R}^{n \times d}$ with $\+x_i^* \in \mathbb{R}^d$, where $d\ll n$ is the latent dimension that can change with $n$. For the data-generating process, the latent space model assumes: 
\begin{equation}\label{eq:likelihood}
    Y_{ij} \sim \mbox{Bernoulli}(\mbox{logistic}(\beta^*+\+x_i^{*'}\+x_j^{*})), \quad i<j \in [n]:=\{1,\ldots, n\}.
\end{equation}
The latent class model proposed in \cite{hoff2002latent}, and elaborated in \cite{handcock2007model,NIPS2007_766ebcd5,krivitsky2009representing,ma2020universal}, has been extensively studied. 
%It has been widely used in various applications such as visualization \citep{sewell2015latent}, edge prediction, and clustering \citep{ma2020universal}. 
The model is based on the idea that each node $i$ in a network can be represented by a latent Euclidean vector, $\boldsymbol{x_i}$. The likelihood of an edge between nodes $i$ and $j$, represented by $Y_{ij}$, is entirely determined by the distance or discrepancy between their respective latent coordinates, $d(\boldsymbol{x_i}, \boldsymbol{x_j})$. After accurately estimating the latent variables of each node, the estimated latent representations can provide useful insights into the network's structure; they can be used as node features for post-processing tasks such as node clustering, missing link prediction, and regression analysis.

Suppose we observed the network associated with high-dimensional covariates $\+Z \in \mathbb{R}^{n \times p}$ where $p$ is allowed to be greater than $n$. $\+Z=[\+z_1,...,\+z_n]'$ with $\+z_i \in \mathbb{R}^{p}$. For example, in a protein-protein interaction network, the covariate information is each protein's characteristics.   We consider the following model: there exists a low-rank linear coefficients matrix of the covariates, denoted by $\+B^* \in \mathbb{R}^{p \times d}$. $\+B^*$ is a matrix consisting of $p$ column vectors, $\+b^*_1,...,\+b^*_p$, where each vector $\+b^*_j \in \mathbb{R}^{d}$. We establish a relationship between the linear transformation of covariates $\+Z\+B^*$ and the latent positions matrix $\+X^*$. For instance, when row vectors of $\+Z$ belong to some different clusters, then the relation $\+X^{*} =\+Z\+B^*$ implies that $\+X^*$ enjoy the same cluster configuration from $\+Z$, as $\+Z\+B^*$ keeps the row cluster configuration of $\+Z$: $\+z_i=\+z_j$ implies $\+B^{*'}\+z_i = \+B^{*'}\+z_j$. To capture the heterogeneous structures discussed previously, we impose specific constraints  on both  the linear coefficients $\+B^*$ and the differences between the true latent matrix and the transformed covariate $\+X^*-\+Z\+B^*$.
 
We first take into account the sparsity of the selection of the high-dimensional covariates.

\begin{equation*}
    \sum_{j=1}^p \ind\{\+b^*_j \ne \+0_d\} \leq s_b, 
\end{equation*}
where $\+0_d$ is a $d$-dimensional vector with all zero components. The constraint implies that only a subset of covariates, $s_b$, is relevant with respect to the latent positions, which is sensible for ultra-high dimensional covariates.  

In addition, to capture the dependence between $\+X^*$ and $\+Z\+B^*$ in a robust fashion, we consider the structure that the differences between the latent positions and the transformed covariates are sparse:

\begin{equation*}
    \|\+X^{*}-\+Z\+B^*\|_{2,0} =\sum_{i=1}^n \ind\{\+x_i^* \ne \+B^{*'}\+z_i\} \leq s_x, 
\end{equation*}
where the norm $\|\cdot\|_{2,0}$ is the number of nonzero row vectors of the matrix. The above two constraint provide the dependence between $\+X^*$ and $\+Z\+B^*$ in a robust fashion:  
\begin{itemize}

    \item $s_x = 0$:  $\+X^{*} =\+Z\+B^*$; which is the traditional case that all the latent positions must follow the exact information obtained from $\+Z$. This reduces to the case that the network and covariates share the same latent factors (e.g., \cite{zhang2022joint}). 
 
    \item $s_b = 0$:  when the additional covariates $\+Z$ are totally independent of the latent positions $\+X^*$, our proposed model can handle the case as it corresponds to $s_b=0$ and $\+B^*=0$ and is included in the structure we considered. This reduces to the case of a latent space model for a network without covariate information (e.g., \cite{hoff2002latent,ma2020universal}).
 
    \item $s_x \ne 0$ and $s_b \ne 0$ but $s_x+s_b\ll n$:
The $s_b$ features in the covariates will contribute to the latent vector in the network and we allow $s_x$ nodes that refuse to receive information from the covariates. For these nodes, the latent positions will only be learned from the network itself. When $\+Z$ has a clustered configuration in the row vectors, the above constraint implies that some nodes may not have the exact cluster configuration as they are in $\+Z$, as illustrated in Figure~\ref{fig:concept}.
\end{itemize}

To estimate parameters $\+X$ and $\+B$ while capturing the proposed structures of the truth, proper priors should be introduced to ensure sparsity on parameters $\{\+b_j\}_{j \in p}$ and $\{\+x_i-\+B'\+z_i \}_{i \in n}$. Here, by the tradeoff between computation and theoretical analysis, we utilize the horseshoe prior proposed in \cite{carvalho2010horseshoe}, but other types of sparsity-induced priors (e.g., spike and slab, \citep{mitchell1988bayesian}, \citep{rovckova2018spike}) may also be effective. Therefore, we consider the following priors:

\begin{align}\label{eq:prior}
\begin{aligned}
    \+b_{j} &\sim \mathcal{N}(\+0, \lambda_{b_j}^2\tau_{b}^2),
        &\tau_{b} \sim \mbox{Ca}^+(0,1), \lambda_{b_j} \sim \mbox{Ca}^+(0,1), \quad j \in [p], \\
    \+x_i &\sim  \mathcal{N}(\+B' \+z_i ,\lambda_{x_i}^2 \tau_x^2),
        &\tau_{x} \sim \mbox{Ca}^+(0,1), \lambda_{x_i} \sim \mbox{Ca}^+(0,1), \quad i \in [n],  
        \end{aligned}
\end{align}
where $\mbox{Ca}^+(0,1)$ stands for half-Cauthy distribution as in \citep{carvalho2010horseshoe}.

We refer to the above prior~\eqref{eq:prior} together with the likelihood~\eqref{eq:likelihood} as the Covariate-assisted Latent Space Model (CALSM). The hierarchical prior consists of two layers: The first layer is the covariate layer, where $\+B$ is a set of latent variables that do not directly influence the likelihood. Its influence on the likelihood occurs only through its possible proximity to the latent positions that can be learned from the information in $\+Z$. Given the latent positions $\+X$, the problem of the covariate layer becomes multivariate sparse regression problems. 
The second layer is the latent space layer, where the information flow between $\+X$ and $\+Z\+B$ is evaluated using another global-local shrinkage prior. If $\+x_i$ and $\+B'\+z_i$ are close enough, then the difference between them is shrunk towards zero, implying node $i$ fully receives information from $\+z_i$. However, if $\|\+x_i-\+B'\+z_i\|_2$ is large enough, then the heavy tail of the global-local shrinkage prior implies that $\+x_i$ can be fully learned through the likelihood information of $\+Y$ and can be far away from $\+B'\+z_i$. 
% \begin{figure}
% \centering
% 			\begin{tikzpicture}[main/.style = {draw,rectangle}] 
% 			\node[main] (1) at (0,0) {$\+x_{i}$};
% 			\node[obs] (3) at (1,1) {$Y_{ij}$};
%                 \node[main] (2) at (2,0) {$\+x_{j}$};
% 			\node[main] (6) at (1,-1) {$\+B$};
%                 \node[main] (4) at (0,-1) {$\+z_i$};
%                 \node[main] (5) at (2,-1) {$\+z_j$};
% 			\plate {plate1} {(1)(2)(3)(4)(5)(6)} {$i, j=1,...,n$};
% 			\edge [-] {1,2}{3}
% 			\edge [dotted] {6}{1,2}
%                 \edge [dotted] {4}{1}
%                 \edge [dotted] {5}{2}
% 			\end{tikzpicture} 
% 		\caption{ Graph representation for CALSM, where the dashed line represents adaptive dependence that may or may not exist.}\label{fig:Aim1}
% 	\end{figure} 
Note that there is no need to estimate $\+B^*$ accurately as it only serves as a latent layer to introduce the dependence between $\+Z$ and $\+X$ and will be marginalized in the posterior and the only target of the proposed prior is to recover the connecting probabilities.

\section{Theoretical Results}\label{sec:theory}

In this section, we aim to provide theoretical support in the homogeneous sparse network setting for the proposed methodology. We begin by identifying a suitable parameter space that reflects our proposed structural constraints. We then list and discuss the assumptions necessary for our analysis. In Theorem~\ref{thm:posterior}, we demonstrate that the rate of contraction of the fractional posterior, a variant of the usual posterior, can converge to the truth with a specific rate. Then in Theorem~\ref{thm:posterior_latent_vector}, we show the convergence in terms of latent vectors under some assumptions of identifiability. Finally, in Theorem~\ref{thm:single_cluster}, we derive the error rates for community detection when $K$-means is applied to the latent vectors estimated. The proofs of all theorems are provided in the Appendix.

We use the fractional posterior framework \citep{walker2001bayesian}, which involves raising the usual likelihood $P(\+ Y\mid \+ X,\beta)$ to a power $\alpha \in (0, 1)$. This creates a pseudo-likelihood $P_\alpha(\+ Y \mid \+ X, \beta)$, which leads to a fractional posterior $P_\alpha(\+ X, \beta\mid \+ Y) \propto P_\alpha(\+ Y \mid \+ X, \beta) \, p(\+ X ) p(\beta)$. Theoretical analysis of fractional posteriors has gained prominence recently \citep{bhattacharya2019bayesian,martin2020empirical,jeong2021posterior}. The convergence of fractional powers ($\alpha < 1$) requires fewer conditions than the usual posterior ($\alpha=1$). Furthermore, the optimal convergence of the fractional posterior directly implies rate-optimal point estimators constructed from the fractional posterior, similar to the usual posterior.

We consider the following homogeneous structured sparse pairwise latent space model as the data-generating process.
\begin{equation} 
\begin{aligned}
\label{likelihood:beta_unknwon}
    	Y_{ij} \stackrel{ind.}\sim  \mbox{Bernoulli}\left[1/\{1+\exp(-\beta^*-\+x_{i}^{*'}\+x_{j}^{*})\}\right], \, i\leq j \in [n], \,\mbox{s.t.,} \\ \quad \|\+X^{*}-\+Z\+B^*\|_{2,0}=\sum_{i=1}^n \ind\{\+x_i^* \ne \+B^{*'}\+z_i\} \leq s_x, \quad   \sum_{j=1}^p \ind\{\+b^*_j \ne \+0_d\} \leq s_b.
     \end{aligned}
\end{equation}
 where the parameter $\beta^*$ controls the sparsity level of  the network and is not known and needs to be estimated. For simplicity of theoretical analysis, we also assume that %the latent dimension $d$ is fixed and does not change with $n$, and 
 the diagonal elements of $\+Y$ are observed. We consider the following assumptions.
 \begin{assumption}[Magnitude of the true parameters] \label{asm:bound} Suppose all the $\ell_\infty$ norm of  latent positions $\+x^*_{i}$ and all latent coefficients $\+b_j$ satisfy $$\max_{i=1}^n\{\|\+x^*_{i}\|_\infty\} \leq C_1, \quad \max_{j=1}^p \|\+b_j^*\|_\infty =O(p^{C_2} ) $$ and 
 for some constants $C_1,C_2>0$.
 \end{assumption}

 \begin{assumption}[Restricted eigenvalues]\label{asm:restricted_eig} There exist constants $\underline{c}, \overline{c}>0$ such that $\forall\, \tilde{\+B} \in \mathbb{R}^{p \times d}$ with $\|\tilde{\+B}\|_{2,1} \lesssim s_b d \log(np)$, we have 
  $$\underline{c} n  \|\tilde{\+B}\|_F^2   \leq \|\+Z\tilde{\+B}\|^2_F \leq \overline{c} n\|\tilde{\+B}\|_F^2.$$
\end{assumption} 
Denote $$\epsilon_{n} = M_0  \left\{\sqrt{ \frac{(1+s_b+s_x)  d\log (np)}{n^2}} \right\}$$ for some constant $M_0>0$.
  \begin{assumption}[Sparsity level]\label{asm:sparsity_level} We assume that the sparsity level of the connected probabilities satisfies
$
    e^{\beta^*} \gg \epsilon_{n}^2. $

\end{assumption}

For the prior setting,  we  have the following assumptions on the sparsity parameter $\beta$:
 \begin{assumption}[Prior for the sparsity parameter]\label{asm:sparsity_prior}  We use prior $ \Pi(\beta)$ such that for $c_4>0$, we have
$$
\Pi\left( |\beta  - \beta^*| \leq  c_4 \epsilon_{n}\right) \gtrsim  e^{-n^2 \epsilon_{n}^2}.
$$
\end{assumption}

Assumption~\ref{asm:bound} is a critical assumption for the homogeneous latent space model. It indicates that all the latent positions are bounded, and the sparsity level of the connections only depends on the parameter $\beta^*$. Assumption~\ref{asm:restricted_eig} is essentially the restricted eigenvalue condition \citep{bickel2009simultaneous} for high-dimensional multivariate regression: when the matrix $\tilde{\+B}$ is close to zero, the behavior of $\+Z$ is similar to that of constant eigenvalues. In this case, we assume that $\+Z$ is already normalized, so there is no need to introduce another factor of $1/n$. In practical applications, we can always normalize the row vectors of $\+Z$ before using the proposed algorithm.  The literature commonly considers Assumption~\ref{asm:sparsity_level} as the smallest sparsity level for which the estimated inner product plus intercept can be consistently recovered, and the community detection error can go to zero. For a network model without covariates, this sparsity level is $\log(n)/n$. The rate $\epsilon_{n}^2$ has improvement over $\log(n)/n$ as long as $d(s_b+s_x) \ll n$ and $\log p =O(\log n)$, which shows the gain of the proposed structure. Assumption~\ref{asm:sparsity_prior} means that the prior $\Pi(\beta)$  has a sufficient mass around $\beta^*$ for $\beta^*$ goes to negative infinity at a specific rate. The assumption can be satisfied by the commonly used normal prior with mean zero with a scale $O(\sqrt{\log n})$, which is proved in Lemma~\ref{lem:scale} in the Appendix.

The following theorem shows that the proposed posterior can converge at an error rate mentioned above:
       \begin{theorem}[Posterior contraction for  connecting probability recovery]\label{thm:posterior}
                Suppose the data generating process follows equation~\eqref{likelihood:beta_unknwon}. Assume that $\epsilon_n \rightarrow 0$ for $n \rightarrow \infty$ and $\log p = O(\log n)$.  Then under the proposed prior~\eqref{eq:prior}, 
                for large enough $n$, any $D\geq 2$ and $\eta>0$, with probability at least $1-2/\{(D-1+\eta)^2n^2 \epsilon_{n}^2\}$, we have 
                \begin{equation*}
    \Pi_{\alpha}\left(\frac{1}{n^2} D_{\alpha}\left( p_{\m X,\beta},  p_{\m X^*,\beta^*}\right) \geq  \frac{D+3\eta}{1-\alpha} \epsilon_{n}^2 \mid \m Y\right) \rightarrow 0.
\end{equation*}
        \end{theorem}
     The above theorem establishes the posterior convergence rate for link connection probabilities, as measured by  $\alpha$-R\'{e}nyi divergence.   The rate $\epsilon_n^2$   has improvement over $d\log(n)/n$ as long as $s_b+s_x \ll n$, which proves the advantage of the model over directly incorporating the node covariates into the logistic link functions  (e.g., Theorem 1 in \cite{yan2019statistical}), as strong regularization can enhance the generalization of the model. In addition, as long as $d s_b = O(n)$, the error rate is at most $d \log (n)/n$. This demonstrates the robustness of the proposed approach: If the covariate information is helpful, then the proposed prior improves the estimation rate based on the covariate information. However, if the covariate information is not useful, then the estimation error rate will be the same as if we only relied on the networks themselves without considering the additional covariate.
The assumptions in Theorem~\ref{thm:posterior} are minimal because they do not require the signal of the relevant covariates to be strong enough for exact selection, as the sparsity of the covariates exists in the latent layer. We only need the prediction $\|\+Z\+B-\+X\|_F$ to be sufficiently small, which requires fewer conditions to be satisfied.

Given the convergence of the connecting probabilities in Theorem~\ref{thm:posterior}, the next step is to derive the theory for the convergence of the latent positions. To achieve that, we need additional assumptions so that the identifiability between $\+X^*$ and $\beta^*$. We adopt the following conditions:

\begin{assumption}[Identifiability]\label{asm:identifiability}
    Assume true latent positions satisfy $\sum_{i } \+x_i^{*} = \+0_d$.
\end{assumption}

\begin{assumption}[Projection]\label{asm:projection}
Consider the event $ B_p = \{  \|\+x_{i}\|_\infty \leq C_5 , \forall  i \in [n],  \|\sum_{i=1}^n \+x_i\|_\infty\leq C_6\} $ for a large enough constant $C_5, C_6>0$. We consider the prior restricted on event  $B_p(\m X), \,\, \tilde \Pi := \Pi(\cdot \cap B_p(\m X))/\Pi(B_p(\m X))$,
to replace the original prior such that $\tilde \Pi(B_p^c)=0$. Without ambiguity, we still use $\Pi(\cdot)$ to denote $\tilde \Pi()$.
\end{assumption}
Asumption~\ref{asm:identifiability} indicates the indetifibility between $\+X^*$ and $\beta^*$. Under this assumption, $\+\beta^*$ can be interpreted as the mean of log odd of all connecting probabilities.   The Assumption~\ref{asm:projection} guarantees that 1) the posterior estimated latent positions are bounded so that the sparsity level of the networks is only explained by $\beta$;  2) the posterior estimated latent positions are roughly centered around the origin. The assumption can be satisfied by adopting the projection into the desired area as proposed in \cite{ma2020universal}. However, the assumption is used solely to simplify proving the theorem and is not utilized in the algorithm. Adopting such a prior will only result in a negligible difference compared to using the original prior  (e.g., see a similar phenomenon in remark 2 of \cite{ma2020universal}).  Then, we have the following theorem about the fractional posterior convergence in $\ell_2$ loss functions.
 \begin{corollary}[Posterior contraction for recovery of inner-products]\label{thm:posterior_latent_vector}
Under the assumptions of Theorem~\ref{thm:posterior},
 if Assumptions~\ref{asm:identifiability} and \ref{asm:projection}  also hold, we  have for large enough constant $M>0$,
                \begin{equation*}\label{eq:result3}
                        \Pi_{\alpha}\left\{\frac{1}{n^2} \sum_{i,j=1}^{n}(\+x_{i}'\+x_{j}-\+x^{*'}_{i}\+x^{*}_{j})^2+(\beta-\beta^*)^2 \geq M\frac{D+3\eta}{1-\alpha}  e^{-\beta^*}\epsilon_{n}^2\mid \m Y\right\} \leq e^{-\eta n^2\epsilon_{n}^2}.
                \end{equation*}
        \end{corollary}

There is an $e^{-\beta^*}$ multiplier for the convergence of the inner product $x'_{i}x_{j}$ with respect to the measure $D_{\alpha}(p_{X,\beta}, p_{X^*,\beta^*})$, as discussed in the related literature \citep{zhang2022directed}. This is due to the weak identifiability of the logistic link functions. When the connected probabilities are close to zero, even if the probabilities can be estimated accurately, the inner product plus the intercept parameter will suffer from an additional error factor determined by the sparsity level of the probabilities due to the flat curvature of the logistic function at probabilities near zero.

Once the latent vectors have been obtained, $K$-means can be used to cluster the subjects. 
For any membership matrix $\+\Xi$, the cluster
	membership of a subject $i$ is denoted by $g_i \in \{1, . . . , K\}$, which satisfies $\Xi_{i g_i}=1$. Let $G_k(\Xi)=\{1\leq i \leq n: g_i=k\}$. We consider the following loss function to evaluate clustering accuracy:
$
	L(\hat{\+\Xi},\+\Xi) =  \min_{\+J \in E_K} \|\hat{\+\Xi} \+J-\+\Xi \|_{2,0},
$
	which represents the number of misclustered subjects, where $E_K$ is the set of all $K \times K$ permutation matrix. Our next result concerns performing community detection for the network while improving the misclustering rate based on the covariate information.
	
	\begin{theorem}[Community detection]\label{thm:single_cluster}
		Suppose the assumptions in Theorem~\ref{thm:posterior_latent_vector} hold. In addition, assume that for the truth $\+X^*$ satisfies:
		\begin{enumerate}
			\item [(1)] $\+X^*$ has $K$ distinct rows: $\+X^*= \+\Xi^* \+X$,  $\+\Xi^*\in \mathbb{M}_{n,K}$ for some $K>0$ and $\+X^* \in \mathbb{R}^{K \times d}$ is full column rank;
			\item [(2)] The smallest singular value of $\+X^*$, denoted as $\lambda$, satisfies $\lambda = \Omega(\sqrt{n})$;
			\item [(3)] Cluster separation: $\delta \leq \min_{i,j} \|\+x^*_{i}-\+x^*_{j}\|_2 $ for any $\+x^*_{i} \ne \+x^*_{j}$, where $\delta>0$;
			\item [(4)] The minimal block size satisfies $\min_{i} |G_i(\+\Xi^*)| = \omega( n \epsilon_n^2 e^{-\beta^*} /\delta^2)$.
		\end{enumerate}	
  Then after performing $K$-means for Bayes estimator $\hat{\+X}$ to obtain estimation of membership matrix $\hat{\+\Xi}$, as $n \rightarrow \infty$, we have
		$$
		 L(\hat{\+\Xi},\+\Xi^*) \lesssim \frac{e^{-\beta^*}\epsilon_n^2}{\delta^2}.
		$$
	\end{theorem}
	
	 The above theorem implies that as long as $\delta =\omega( e^{-\beta^*/2} \epsilon_n)$, a consistency of community detection can be achieved for large enough $n$, where the misclustering error can converge to $0$. This demonstrates the advantage of absorbing information from the additional covariates over without covariates in community detection, whose error rate is $e^{-\beta^*}d\log n/(n\delta^2)$ (e.g., see Theorem 2 in \cite{zhang2022directed}). Furthermore, all the above Theorems hold true for sparse networks with link probabilities tending to $0$ at an order no faster than $\epsilon_n^2$ or the expected node degree growing at the order of $(1+s_b+s_x) d\log(np)/n$.  Regarding assumptions in Theorem~\ref{thm:single_cluster}: Assumption (1) is prevalent in the literature on community detection for networks. Assumption (2) requires the lower bound of the order $\sqrt{n}$ for the latent matrix, which holds for an $n \times d$ random matrix whose entries are i.i.d. \citep{vershynin2010introduction}.
  Assumption (3) represents a minimum separation between different clusters. In the extreme case where both $s_x$ and $s_b$ are constants, the requirement of minimal separation of the latent vectors is only at $1/n$, instead of $1/\sqrt{n}$ in the literature.   Finally, following \cite{lei2015consistency}, the $K$-means can be easily replaced by some variants that are easier to use, such as $(1+\epsilon)$ $K$-means and $K$-medians.

%Note that the theoretical results proposed are within the framework of homogeneous sparse network settings. The extension to heterogeneous sparse networks is left for future research.

\section{Computation}\label{sec:computation}
Sampling-based approximation algorithms become computationally inefficient for networks with many nodes and high-dimensional covariates. Instead, using variational inference to accelerate computation for latent space models for networks has recently gained attention, e.g., \cite{loyal2023eigenmodel,zhao2022factorized,zhao2022structured}. We develop two complementary variational inference algorithms for posterior approximation under the proposed framework. The first implementation, optimized for accessibility, provides a user-friendly R interface accelerated through the \textit{Rcpp} package \citep{Eddelbuettel2011Rcpp}. The second implementation addresses scalability challenges via  PyTorch, enabling GPU-accelerated inference for networks exceeding $10^4$ nodes. 
\subsection{Coordinate Ascent Variational Inference}
The first algorithm uses closed-form iterations of a coordinate-ascent method,
which relies on the conditionally conjugate nature of the CALSM prior.  We start our variational targets as the fractional
posterior  and aim to find the best approximation (in terms of KL divergence) from a mean-field variational (MF) family
  \begin{equation}\label{eq:MF}
                q(\+X,\beta,\+B) = \left\{\prod_{i=1}^{n}q(\+x_{i}) \right\}\left\{\prod_{j=1}^{p}q(\+b_{j}) \right\}q(\beta). 
        \end{equation}
         The objective function of the variational inference is  
        \begin{align}\label{eq:KL}
                \begin{aligned}
                        \hat{q}(\+X,\beta,\+B) &=\argmin_{q(\+X,\beta,\+B) \in \Gamma} D_{KL}  \left\{ q(\+X,\beta,\+B) \mid \mid p_\alpha(\+X,\beta,\+B \mid \m Y) \right\} \\&=\argmin_{q(\varTheta,\+\beta) \in \Gamma} -\E_q \left\{\log \left(\frac{p_\alpha(\m Y,\+X,\beta,\+B)}{q(\+X,\beta,\+B)}\right)\right\},
                \end{aligned} 
        \end{align}
        where the term $\E_q\{\log(p_\alpha(\m Y,\+X,\beta,\+B)/q(\+X,\beta,\+B))\}$ is the evidence-lower bound (ELBO) and  $p_\alpha(\+X,\beta,\+B \mid \m Y)$ is the marginal fractional posterior $p_\alpha(\+X,\beta,\+B \mid \m Y)= \int p_\alpha(\+X,\beta,\+B, \+\lambda_x,  \\  \tau_x, \+\lambda_b, \tau_b  \mid \m Y) d  \+\lambda_x d\tau_x d \+\lambda_b d\tau_b$. However, obtaining the marginal variational posterior~\eqref{eq:MF} through marginalization is inefficient.  We use the joint variational posterior to approximate the marginal variation posterior.
        Furthermore, we adopt the variable augmentation that the square of the half-Cauchy distribution can be expressed as a mixture of Inverse-Gamma distributions \citep{neville2014mean}. 
        % \begin{align*}
        %     \lambda_{x_i}^2 \mid v_{x_i} \sim \mbox{IG}(1/2,1/v_{x_i}),\,\, v_{x_i} \sim \mbox{IG}(1/2,1). \,\,
        %      \tau_{x}^2 \mid v_{x} \sim \mbox{IG}(1/2,1/v_{x}),\quad v_{x} \sim \mbox{IG}(1/2,1). \\
        %       \lambda_{b_j}^2 \mid v_{b_j} \sim \mbox{IG}(1/2,1/v_{b_j}),\,\,\quad v_{b_j} \sim \mbox{IG}(1/2,1).\,\,
        %        \tau_{b}^2 \mid v_{b} \sim \mbox{IG}(1/2,1/v_{b}),\quad v_{b} \sim \mbox{IG}(1/2,1),  
        % \end{align*}
        % where $\mbox{IG}(a,b)$ stands for Inverse Gamma distribution with shape parameter $a$ and scale parameter $b$.
        Then the objective of KL minimization is as follows: 
        \begin{equation*} 
                \begin{aligned}
                        \hat{q}(\+X,\beta,\+B, \+\lambda_x, \tau_x, \+\lambda_b, \tau_b  ) 
                        &=\argmin_{q( \+X,\beta,\+B, \+\lambda_x, \tau_x, \+\lambda_b, \tau_b)   \in \Gamma} -\E_q \left\{\log \left(\frac{p_\alpha(\m Y, \+X,\beta,\+B, \+\lambda_x, \tau_x, \+\lambda_b, \tau_b )}{q(\+X,\beta,\+B, \+\lambda_x, \tau_x, \+\lambda_b, \tau_b )}\right)\right\},
                \end{aligned}  
        \end{equation*}
        where now the variational family $\Gamma$ is defined as:
        \begin{align}\label{eq:SMF_joint}
        \begin{aligned}
                q(\+X,\beta,\+B, \+\lambda_x, \tau_x, \+\lambda_b, \tau_b) =   \left\{\prod_{i=1}^{n}q(\+x_{i})q(\lambda_{x_i})q(v_{x_i}) \right\}q(\tau_x)q(v_x) \times\\
                \left\{\prod_{j=1}^{p}q(\+b_{j}) q(\lambda_{b_j})q(v_{b_j})\right\}q(\tau_b)q(v_b)q(\beta).
                        \end{aligned}
        \end{align}
        The variational family defined in equation~\eqref{eq:SMF_joint} allows for updating all the scales in the inverse-Gamma conjugate family.   For any parameter $\theta \in \left\{ [\+x_i]_{i \in [n]},[\lambda_{x_i}]_{i \in [n]}, [v_{x_i}]_{i \in [n]}, \tau_x, v_x, \beta,\right. \\ \left. [\+b_j]_{j \in [p]},  [\lambda_{b_j}]_{j \in [p]}, [v_{b_j}]_{j \in [p]}, \tau_b, v_b \right\}$,
the updated variational posterior under the proposed family~\eqref{eq:SMF_joint} can be obtained through CAVI to maximize the ELBO \citep{blei2017variational}: 
	\begin{align}\label{eq:mf}
	q^{(new)}(\theta) \propto \exp [\E_{-\theta}\{\log p_\alpha(\+X,\beta,\+B, \+\lambda_x, \tau_x, \+\lambda_b, \tau_b)\}],
	\end{align}  where $\E_{-\theta}$ is the expectations taken with respect to the densities without the parameter $\theta$. 	Finally, for the Bernoulli likelihood, we adopt the tangent transform approach by introducing auxiliary variables $\xi_{{ij},{i,j \in [n]}}$ as proposed by \cite{jaakkola2000bayesian}  to obtain closed-form updates. Algorithm~\ref{alm:CAVI} presents the pseudocode for our approach, with detailed update steps available in Section~\ref{sec:CAVI} of the supplementary material.

\begin{algorithm}
\caption{Pseudocode for CAVI algorithm for Covariate-assisted latent space model}\label{alm:CAVI}
\begin{algorithmic}[1]
\State \textbf{Initialize parameters:} Set initial values for $\mu_{\beta_0}$, $\sigma^2_{\beta_0}$, $\bm{\mu}_{x_i}$, $\Sigma_{x_i}$, $\bm{\mu}_{\+{b}_j}$, and $\Sigma_{\+{b}_j}$.

\State \textbf{Update} $\xi^2_{ij}$ for all $i,j$ using tangent transformation by equation~\eqref{update_xi};

\State \textbf{Update} $\mu_{\beta_0}$ and $\sigma^2_{\beta_0}$' by equation~\eqref{update_beta};

\State \textbf{Update} $\bm{\mu}_{x_i}$ and $\Sigma_{x_i}$ for each node $i$ by equation~\eqref{update_X}; 

\State \textbf{Update scales for each nodes} $a_{\lambda_{x_i}}, b_{\lambda_{x_i}}, a_{v_{x_i}}, b_{v_{x_i}}, a_{\tau_x}, b_{\tau_x},a_{v_x}, b_{v_x}$,  by equation~\eqref{update_scale_X};

\State \textbf{Update} $\bm{\mu}_{\+{b}_j}$  $\Sigma_{\+{b}_j}$ for each covaraite variable $k$ by equation~\eqref{update_B};

\State \textbf{Update scales for covarite coefficients} $a_{\lambda_{b_j}}, b_{\lambda_{b_j}}, a_{v_{b_j}}, b_{v_{b_j}}, a_{\tau_b}, b_{\tau_b},a_{v_b}, b_{v_b}$ by equation~\eqref{update_scale_B};

\Repeat
    \State Perform steps 2-8;
\Until convergence

\end{algorithmic}
\end{algorithm}

\subsection{Stochastic Variational Inference}
For large-scale sparse network models, stochastic variational inference (SVI) algorithms have increasingly attracted attention as efficient solutions, as demonstrated in recent work such as \cite{loyal2024fast}. In this paper, we propose a computationally efficient SVI algorithm specifically designed for posterior inference under our proposed CALSM. This algorithm achieves a computational complexity of $O(|E|)$ per iteration, where $|E|$ represents the number of positive edges.

We approximate the posterior using a factorized variational family:  
$
q(\boldsymbol{X}, \beta, \boldsymbol{B}, \lambda_x, \tau_x, \lambda_b, \tau_b) = q(\beta)q(\boldsymbol{X})q(\boldsymbol{B})q(\lambda_x)q(\tau_x)q(\lambda_b)q(\tau_b),
$
with Gaussian distributions for continuous parameters:  
$
q(\beta) = \mathcal{N}(\mu_\beta, \sigma_\beta^2), \quad  
q(\boldsymbol{X}) = \mathcal{N}(\boldsymbol{\mu}_X, \sigma_X^2\boldsymbol{I}), \quad  
q(\boldsymbol{B}) = \mathcal{N}(\boldsymbol{\mu}_B, \sigma_B^2\boldsymbol{I}),
$
and Gamma variational distributions for scale parameters: $ 
q(\lambda_x) = \text{Gamma}(\alpha_{\lambda_x}, \beta_{\lambda_x}), \,  
q(\tau_x) = \text{Gamma}(\alpha_{\tau_x}, \beta_{\tau_x})$, $ 
q(\lambda_b) = \text{Gamma}(\alpha_{\lambda_b}, \beta_{\lambda_b}), \, 
q(\tau_b) = \text{Gamma}(\alpha_{\tau_b}, \beta_{\tau_b}).
$ 
The Gamma distributions are used as they have sufficient mass around zero, which can promote near-zero values for the scale parameters. The convergence of the proposed algorithm critically depends on initializing Gamma hyperparameters. Through systematic experimentation, we recommend:  
$
\alpha_{\lambda_x} = \alpha_{\lambda_b} = 10.0, \,\beta_{\lambda_x} = \beta_{\lambda_b} = 10.0; \, 
\alpha_{\tau_x} = \alpha_{\tau_b} = 0.1, \, \beta_{\tau_x} = \beta_{\tau_b} = 1.0.
$
This configuration initializes node-specific scales ($\lambda_x, \lambda_b$) with mean 1.0 and variance 0.1, balancing flexibility for growth or shrinkage. Global scales ($\tau_x, \tau_b$) are initialized with mean 0.1 and variance 0.1 to promote sparsity while allowing adaptation during inference.  

With the specification of the variational family, our SVI approach approximates the expectation of the KL divergence target in equation~\eqref{eq:KL} by employing an unbiased gradient estimate. This is achieved by drawing $S$ samples from the variational distribution and applying Monte Carlo approximation techniques.  To enable inference on large networks, we implement a two-stage edge subsampling strategy:  
 1). Minibatch sampling: Randomly select a subset of observed positive edges $\tilde{E}^+$;
2). Negative edge sampling: For each positive edge $(i,j) \in \tilde{E}^+$, sample $5$ negative edges $(i,j')$, where the number $5$ is empirically optimal for efficiency- accuracy trade-offs.
Then the log-likelihood is approximated via an unbiased weighted estimator:  
\begin{equation*}
\log p(\boldsymbol{Y}|\boldsymbol{X},\beta) \approx \frac{|E^+|}{|\tilde{E}^+|} \sum_{(i,j) \in \tilde{E}^+} \log p(Y_{ij}=1|\boldsymbol{X},\beta)
+ \frac{|E^-|}{|\tilde{E}^-|} \sum_{(i,j) \in \tilde{E}^-} \log p(Y_{ij}=0|\boldsymbol{X},\beta),
\end{equation*}
where $|E^+|, |E^-|$ denote total positive/negative edges. This reduces computational complexity of the likelihood from $O(n^2)$ to $O(|E^+|)$, enabling application to sparse networks with a large number of nodes. The implementation details of the proposed algorithm are provided in the real data analysis of  Section~\ref{sec:data_analysis} and the supplementary materials.

\section{Data Analysis}\label{sec:data_analysis}

In this section, we demonstrate the advantages of our proposed model through data analysis. We start by conducting simulations to compare our approach with other related approaches in terms of connecting probability recovery and community detection. The simulation settings have been carefully designed to cover three scenarios: (1) weak network signals where additional covariate information is valuable, (2) strong network signals where covariate information adds minimal value, and (3) mismatches between network and covariate information that render covariates not useful. In real data analysis, we evaluate the community detection performance by comparing it to other methods using different datasets.

\subsection{Simulations}\label{sec:simu}

\begin{figure}
    \centering
    \includegraphics[width=\linewidth]{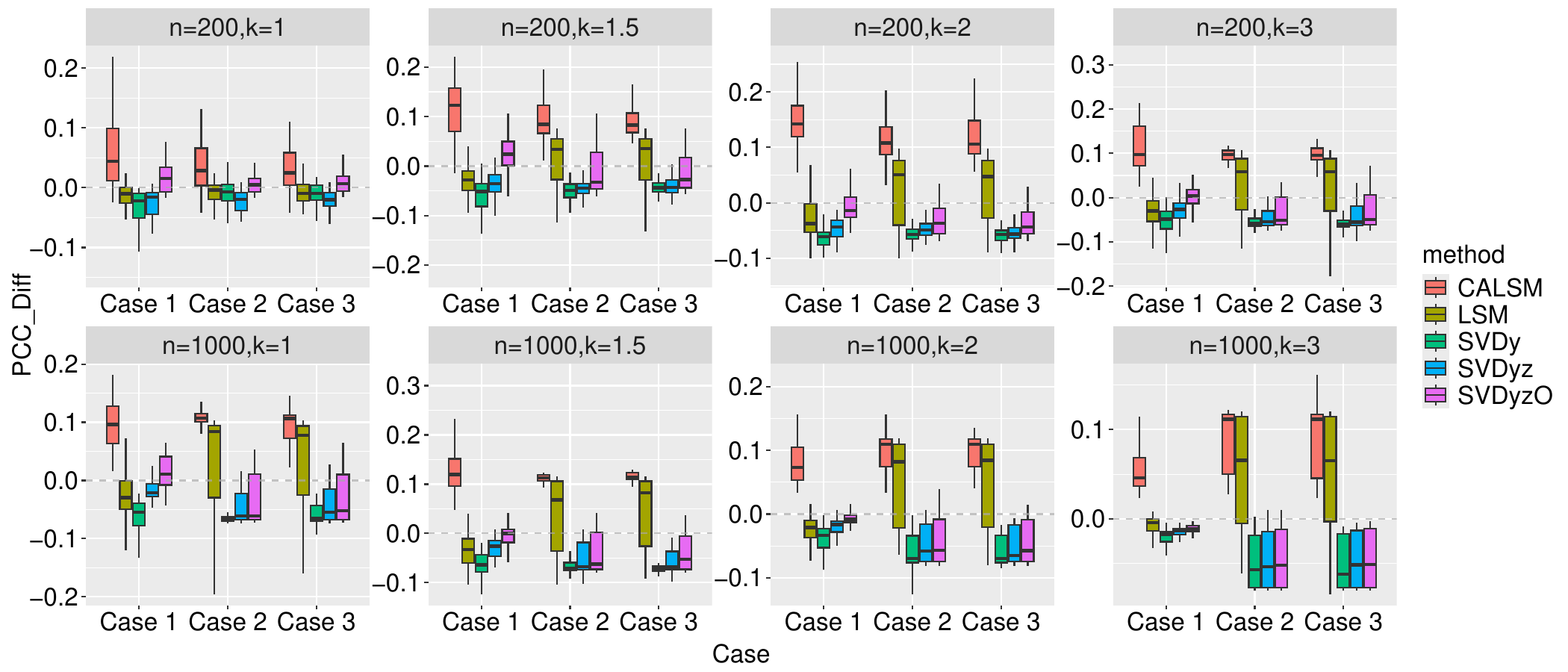}
    \caption{Comparison of community detection methods in terms of $\text{PCC\_Diff}$, which is the PCC of each method minus the average PCC across all methods. The subfigures show the performances for different sample sizes ($n$) and signal strength ($k$) across three different cases. A higher $\text{PCC\_Diff}$ indicates better performance. In weak signal and well-specified covariate settings, CALSM performs the best, and SVD performs the second best with using accurate information about non-zero covariate variable sets of node features. In strong signal or misspecified covariate settings, CALSM also performs best while the latent space model using only network information performs the second best.}
    \label{fig:PCC}
\end{figure}

 \textbf{Connecting probability recovery:} We conducted simulations to compare different methods in terms of connecting probability recovery. We compared our method with the following cases: 1) latent space model with no covariate information (LSM, \cite{ma2020universal}) implemented in R package \textit{randnet} \citep{Li2023randnet}; 2) SVDy, perform SVD to obtain the best rank $d$ approximation for $\+Y$, which only uses the network information; 3) SVDyz, combine the observed network and the covariates $[\+Y,\tilde{\+Z}]$ to perform a rank $d$ approximation and the estimator $\hat{\+Y}$ is the corresponding sub-matrix of the low-rank estimator, where $\tilde{\+Z}= \+Z/(\max_{i,j}|Z_{ij}|)$ to make sure network $\+Y$ and additional covariate $\tilde{\+Z}$ are in the same scales; 4) SVDyzO, the same setting as  SVDyz but only uses the exact true set of variables of $\+Z$. The covariates $\+{Z}$ with $n=200,1000$ samples and $p=500$ features are created by generating multivariate normal samples based on mean zero and the covariance matrix $ \+I_{p \times p}$.  We generate a $p \times d$ matrix $\+B^*$ by randomly assigning $5$ rows with values from the set $\{-2,2,-1.5,1.5\}$ and setting the rest to zero. We then generate an intermediate value $\tilde{\+X}$ in three different settings: 
\begin{itemize}
    \item Case 1. Set $\tilde{\+X} = \+Z\+B^*$.
    \item Case 2. Select $5$ random rows from $\tilde{\+X}$ in Case 1 and replace all the elements with random uniform samples from the interval $[-2,2]$.
    \item Case 3. Replace all rows of $\tilde{\+X}$ in Case 1 with random uniform samples from the interval $[-2,2]$ to ensure that the network has no relation with the covariates.
\end{itemize}
Roughly speaking, Case 1 represents well-specified covariate settings where additional node information uniformly contributes to generating the networks. Case 2 corresponds to scenarios with partial node-level mismatches, where covariate information is not useful for some nodes. Case 3 exemplifies completely misspecified covariate settings where all additional node information lacks utility. To systematically vary the separation between latent positions, we generate $\+X^* = \tilde{\+X} /(max_{i,j}|\tilde{X}_{i,j}|)\times k$ with $k \in \{1,1.5,2,3\}$. The value of $k$ plays a crucial role in determining the relative contribution of network signals versus covariate information. When $k$ is small, network signals remain weak, making accurate incorporation of covariate information essential for optimal performance. Conversely, with large $k$ values, strong network signals dominate, minimizing the influence of additional covariates and elevating correct model specification as the primary determinant of performance quality. We assign the values of $\beta^*= -2$ to control the overall sparsity level of the networks (the mean connecting probability is around $1/(1+\exp(2)) \approx 0.1$), then generate the networks by equation~\eqref{eq:likelihood} for each of the above cases. 
We use the CAVI algorithm with a stopping criterion of either 200 cycles or a mean difference of less than $10^{-4}$ between the estimated denoised connecting probabilities in two consecutive cycles. We use  Pearson's correlation coefficient (PCC) to measure the discrepancy between true and estimated probabilities. To perform a fair comparison for different approaches, we report $\text{PCC\_Diff}$, which is the PCC of each method minus the average PCC across all methods for the specific generated data. The simulation of each case is repeated $25$ times. The simulation results in Figure~\ref{fig:PCC} demonstrate our theoretical results on the adaptivity and robustness of the proposed method. 

In weak signal settings (with $n=200$ and $k=1,1.5,2, n=1000$  for Case 1), method performance critically depends on effectively incorporating useful covariate information. Our approach demonstrates superior performance by employing a shrinkage prior that enables the implicit selection of informative components within node covariates. SVDyzO achieves the second-best performance by utilizing the exact true set of variables, but its performance decreases significantly when providing mismatched information between networks and covariates. Conversely, in strong signal settings (with $k=1.5,2,3$ for Cases 1 and 2) and with misspecified covariates (Case 3), performance superiority depends on model specification correctness and robustness against misleading covariate information. In these scenarios, CALSM again demonstrates optimal performance through its shrinkage prior mechanism that accommodates potential mismatch between network and covariate information. LSM performs second-best in these contexts precisely because it only uses network information. Notably, across all experimental cases and parameter settings, our approach consistently outperforms all comparative methods, validating its adaptivity to varying signal strengths and robustness to covariate misspecification.

\begin{figure}
    \centering
    \includegraphics[width=\linewidth]{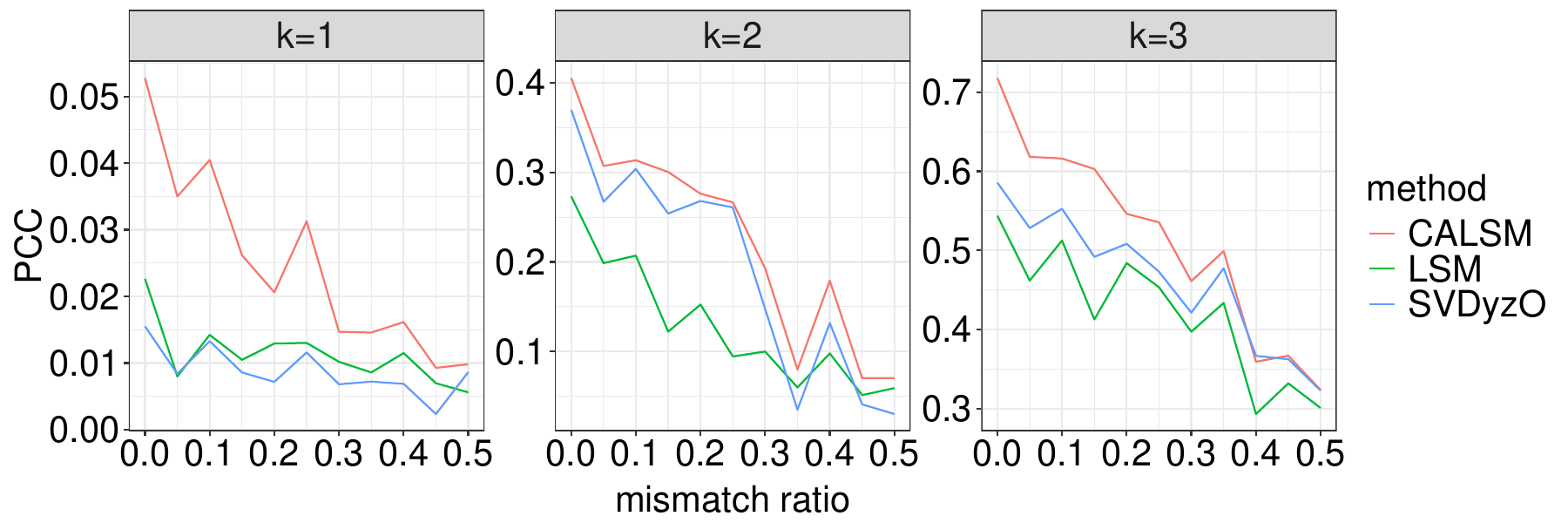}
    \caption{ Comparison of median PCC across $50$ Monte Carlo simulations for different approaches, mismatch ratios, and signal strengths. As the mismatch ratio increases to $0.5$, the performance of  CALSM  will still be better than LSM, while when the mismatch ratio is close to zero, CALSM can benefit a lot from absorbing the additional covariates information similar to joint SVD with Oracle index information of the additional covariates.}
    \label{fig:mismatch}
\end{figure}

We also conducted a simulation to compare the performance of CALSM, LSM and SVDyzO for various mismatch ratios from $0$ to $0.5$, the number of mismatched nodes divided by the number of nodes $n=200$, and different signal strengths $k$. In the simulation, we followed Case 2 settings and changed the process of selecting rows from $\tilde{\+X}$: Instead of selecting 5 random rows, we selected $n$ multiplied by the mismatching ratio rows and replaced them with random samples from uniform intervals. The rest of the settings remain the same, and the simulation is repeated $50$ times. The simulation results for signal strengths of $k=1,2,3$ are shown in Figure~\ref{fig:mismatch}. From the figure, we observed that as the mismatch ratio increased from $0$ to $0.5$, CALSM's performance remained superior to LSM's. However, when the mismatch ratio approached zero, CALSM showed significant improvement by incorporating additional covariate information, similar to joint SVD with oracle subset information of the additional covariates.

 \textbf{Community detection:}  
In addition, we conduct simulations to demonstrate the effectiveness of CALSM in community detection. First, we generate additional covariates $\+Z$ using a binary distribution with probabilities of $0.5$ for $\{-1,1\}$. Then we create a matrix $\+B^*$ with dimensions $p \times d$ by randomly assigning non-zero values to $4$ rows from the set $\{-1,1\}$ with equal probabilities and setting the rest to zero. Due to all discrete settings, the resulting $\tilde{\+X}^* = \+Z\+B^*$ will yield a maximum of $2^4 = 16$ clusters, and the remaining data generation process is similar to connecting probability recovery for different Cases, $k=1,1.5,2,3$, $d=2,n=200,500$ and $\beta^*=-2$. When generating mismatch rows in Cases 2 and 3, we random permute the selected row vectors so that the community information between $\+Z$ and $\tilde{\+X}$ are inconsistent for these rows.

\begin{figure}
    \centering
    \includegraphics[width=\linewidth]{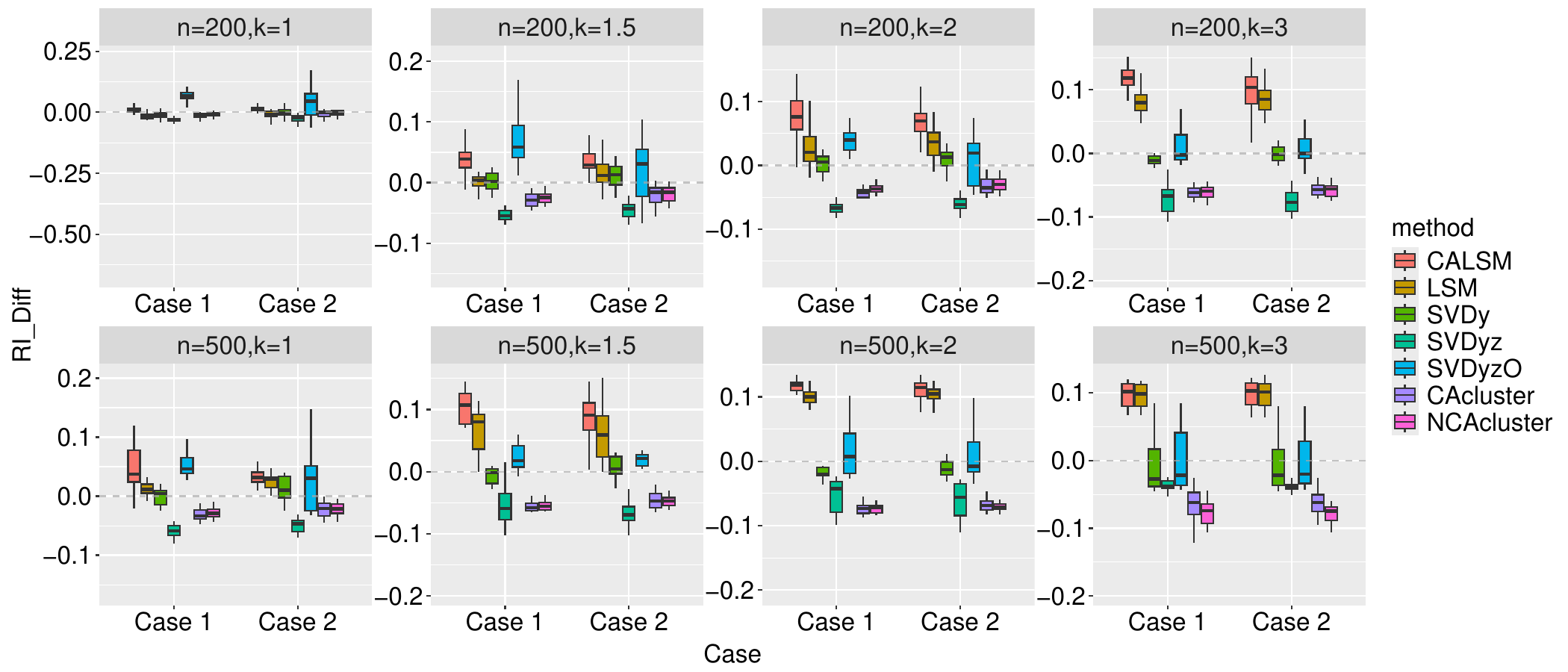}
    \caption{Comparison of community detection methods in terms of $\text{RI\_Diff}$, which is the RI of each method minus the average RI across all methods for Cases 1 and 2. A higher $\text{RI\_Diff}$ indicates better performance. When the signal strength is weak ($k=1$ or $n=200, k=1.5$), CALSM performs second best compared to SVD with oracle information of non-zero covariate variable set, while it performs the best in middle and strong signal settings.}
    \label{fig:community_detection}
\end{figure}

\begin{figure}
    \centering
    \includegraphics[width=\linewidth]{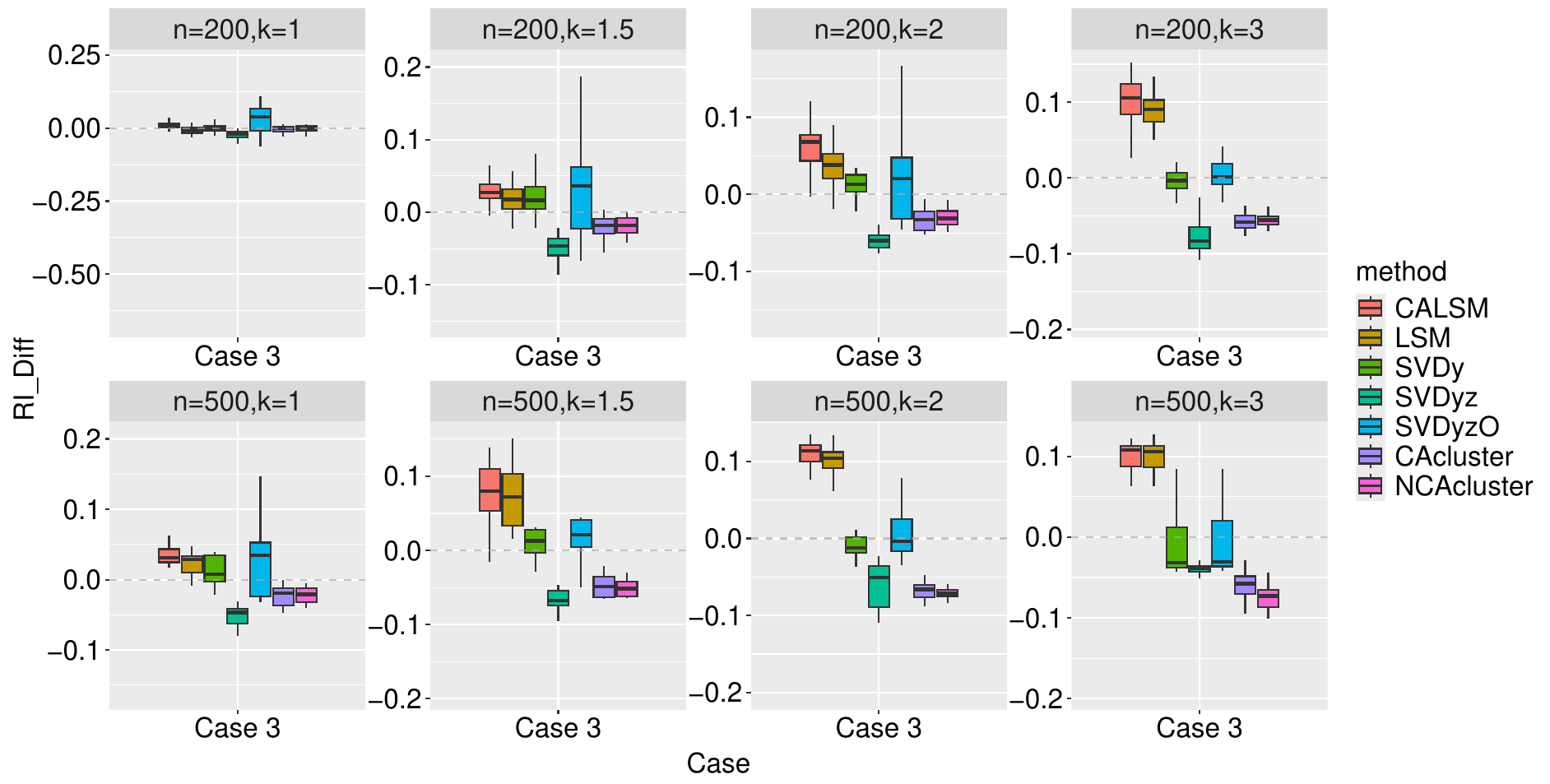}
    \caption{Comparison of community detection methods in terms of $\text{RI\_Diff}$, which is the RI of each method minus the average RI across all methods for Case 3 (misspecified covariates). A higher $\text{RI\_Diff}$ indicates better performance. When the signal strength is weak ($k=1$ or $n=200, k=1.5$), all methods perform similarly poorly. When the signal strength is not weak, CALSM performs the best in the misspecified covariate settings, performing better than or similar to the latent space model that only uses network information.}
    \label{fig:community_detection_2}
\end{figure}

We compare CALSM to all the above algorithms together with CAcluster \citep{binkiewicz2017covariate} and NACcluster \citep{hu2024network}, implemented using R package \textit{NAC}.  We use the CAVI algorithm for our approach, with a stopping criterion of either 500 cycles or a mean difference of less than $10^{-4}$ between the estimated denoised connecting probabilities in two consecutive cycles.  The simulation of each case is repeated with $25$ times. After obtaining the estimate of the latent vectors , we normalized all the latent subjects and then applied $K$-means with known $K$ as the true number of clusters. Finally, the Rand index (RI) between the true cluster and the estimated cluster determined by $K$-means is used to assess the accuracy of the community detection.  Again, we report $\text{RI\_Diff}$, which is the RI of each method minus the average RI across all methods for the specific generated data. The results are shown in Figure~\ref{fig:community_detection} for Cases 1 and 2 and Figure~\ref{fig:community_detection_2} for Case 3. Figure~\ref{fig:community_detection} and Figure~\ref{fig:community_detection_2} reveal several key performance patterns. In extremely weak signal conditions with well-specified covariates, where latent vector separation is difficult to detect ($k=1$ across all Cases and $k=1.5, n=200$ for Cases 1 and 2), CALSM performs second only to SVDyzO, which benefits from oracle knowledge of the non-zero covariate variable set. As signal strength increases ($k=1.5$ with $n=500$, and all Cases for $k=2,3$), CALSM consistently outperforms all competing methods across all cases, demonstrating significantly superior performance. Notably, in completely misspecified covariate scenarios (Case 3) with extremely strong signal strength ($n=500, k=3$), CALSM performs comparably to the latent space model only using network information, with both methods achieving the best performance. Meanwhile, CAcluster and NCAcluster generally perform similarly to SVDyz across most scenarios, as all three methods utilize all covariate information without selection, resulting in suboptimal performance. These comprehensive results validate our proposed method's capacity for effective covariate information selection and robust performance across varying signal strengths and potential mismatch scenarios, which is particularly valuable given that true signal strength and potential mismatches are typically unknown in real-world applications.

\subsection{Community detection for real world networks}\label{sec:real_data}

In this section, we demonstrate the performance of our proposed methodology on three real-world networks with covariate information.
 \textbf{Yeast protein-protein interaction.} The Yeast protein-protein interaction network dataset is a comprehensive compilation of interactions among proteins in yeast, collected by \cite{von2002comparative}, which serves as a critical resource for understanding cellular processes. We utilize the version of the dataset recorded in the R package \textit{igraphdata} \citep{Csardi2015igraphdata}. The dataset consists of 2617 proteins interconnected through 11915 unique interactions. Examples of the protein names include `YLR197W' and `YOR039W'. The proteins belong to 14 categories, with 13 meaningful labels indicating the utility and characteristics of the corresponding proteins, with one label being unavailable. In addition to the network connecting information, there are additional descriptions for each protein, expressed in one sentence. For instance, for the protein `YLR197W', the description is `SIK1 involved in pre-rRNA processing', and for `YOR039W', it is `CKB2 casein kinase II beta chain'. The descriptions are useful in determining the utility and characteristics of the proteins, as proteins with similar descriptions have a high probability of being connected. We construct a word occurrence matrix for additional $1197$ node features, where each predictor is the $0/1$ appearance of a specific word in the corresponding protein description that appears at least twice among all proteins. \textbf{Cora citation.} The Cora citation dataset \citep{mccallum2000automating} is a collection of scientific publications classified into $7$ classes.  The dataset consists of 2708 scientific publications linked by 5429 citations, which are considered undirected in our context. Each publication is featured by a 0/1-valued word vector, indicating the absence or presence of specific words from a dictionary containing 1433 unique words. These word vectors offer a detailed representation of the content of each publication. \textbf{LastFM100.}
The LastFM Asia social network dataset \citep{rozemberczki2020characteristic} is a collection of LastFM users from Asian countries gathered from the public API in March 2020. In this social network, nodes represent LastFM users and edges indicate mutual follower relationships. The features of each user are $0/1$ dummy values based on the artists they like. The main objective is to detect the latent community of users, where the true label is assumed as their country information. We are focusing on a subset of this dataset containing countries with fewer than 100 users. According to the analysis of \cite{hu2024network}, the covariate information for this subset was found to be not useful as a whole. As a result, the community detection method using covariate information, such as NACcluster, may not perform as well as the method that does not use covariate information. We want to use this smaller subset to test the robustness of our method. \textbf{PubMed citation.} The PubMed dataset \citep{namata2012query} contains 19,717 scientific publications related to diabetes, categorized into three distinct classes and interconnected through 44,338 citation links. Each publication is represented by a TF/IDF weighted word vector derived from a dictionary of 500 unique terms. The extreme sparsity of this network, combined with its large-scale node structure, presents significant computational challenges that our SVI algorithm is specifically designed to address.

We compared CALSM to several community detection methods (SVDy, SVDyz, LSM, CAcluster, and NACcluster) across all three datasets while implementing the SVI algorithm. For our SVI implementation, we configured the model with a 5-dimensional latent space ($d = 5$) for Cora, Yeast, and LastFM100 datasets. However, for the PubMed dataset, we used a reduced dimensionality ($d = 3$) since PubMed contains only three distinct clusters. The optimization procedure employed the AdamW optimizer \citep{loshchilov2019decoupled} with an initial learning rate of $\eta = 0.005$ and weight decay coefficient of $10^{-4}$ to mitigate overfitting risk. Network processing utilized batch sizes of 2048 for PubMed, 1024 for Cora and Yeast datasets, and 128 for LastFM100. All other algorithmic parameters remained consistent across datasets to avoid dataset-specific parameter tuning. To ensure robust ELBO approximation, we utilized $S=10$ Monte Carlo samples per gradient step. The training process continued for a maximum of 200 epochs with an early stopping mechanism triggered after 50 consecutive epochs without improvement. To navigate optimization landscape challenges, we implemented a learning rate scheduler that reduced the learning rate by a factor of 0.5 following 20 epochs without improvement. Additionally, gradient clipping with a threshold of 1.0 was applied to prevent exploding gradients. After obtaining latent vector estimates, we normalized all vectors and applied $K$-means clustering with the true number of clusters specified as a known parameter. Community detection accuracy was assessed by calculating the RI between true cluster memberships and the estimated cluster memberships determined by the $K$-means algorithm. The reported results for Cora, Yeast and LastFM100 were obtained on a Windows 11 PC with an Intel(R) Core(TM) Processor (i9-14900 v3@2.00GHz) and 32G RAM, while the results for CALSM for PubMed are implemented with a NVIDIA RTX 4090 GPU.

The characteristics of the  Cora, Yeast and LastFM100 networks and their community detection accuracy are summarized in Table~\ref{tab:real_data} and the community detection accuracy and computation time for PubMed network are summarized in Table~\ref{tab:real_data_2}. The results demonstrate that our proposed CALSM method outperforms all competing approaches for all the datasets. For these datasets, NACcluster and CAcluster achieve better performance than SVDy, with CALSM and NACcluster ranking as the top two methods overall, indicating that covariate information provides substantial information for both networks. In these scenarios, our method effectively leverages the additional information through the selective incorporation of covariate features, leading to superior performance. Conversely, for the LastFM100 dataset, SVDy outperforms both CAcluster and NACcluster, suggesting that the additional covariate information may be less informative or potentially misleading, consistent with findings reported in \cite{hu2024network}. In this challenging scenario, our approach maintains performance comparable to LSM, effectively behaving as if covariate information were disregarded, thus demonstrating the robustness of our proposed model. For the PubMed dataset, LSM could not allocate enough memory due to the extremely large scale of the networks, while the computational time of CALSM was less than one hour, which is only slightly more than CAcluster that tunes $\alpha$ in equation~\eqref{eq:literature} but with a better accuracy. These contrasting cases collectively validate the adaptivity of our model across both covariate-informative and covariate-uninformative scenarios and the scalability of our algorithm, confirming their practical utility for real-world network analysis.
\begin{table}[ht]
    \centering
     \caption{ Data characteristic and community detection results for  Yeast, Cora and LastFM100 datasets. The performance of the method is evaluated using the Rand Index, which measures the similarity between the true community labels and the estimated labels. A higher Rand Index indicates better performance. The top-performing algorithm is shown in bold, while the second top-performing algorithm is in italics.} \label{tab:real_data}
    \begin{tabular}{lccccccccc}
        \toprule
        \multicolumn{4}{c}{Data Description} & \multicolumn{6}{c}{Method Performance} \\
        \cmidrule(lr){1-4} \cmidrule(lr){5-10}
        Data & $n$ & $p$ & $K$ & LSM & CALSM & SVDy & SVDyz & CAcluster & NACcluster \\
        \midrule       
       Cora & 2708 & 1432 & 7 & 0.784 & \textbf{0.846} & 0.670 & 0.742 & 0.751& \textit{0.817} \\
        Yeast & 2617 & 1197 & 13 & 0.813 & \textbf{0.838} & 0.802 & 0.800 & 0.801 & \textit{0.824}  \\
        LastFM100 & 396 & 403 & 7 & \textit{0.882} & \textbf{0.884} & 0.865 & 0.767 & 0.863 & 0.799  \\ 
        \bottomrule
    \end{tabular}
   
\end{table}

\begin{table}[ht]
    \centering
     \caption{RI and computation time comparisons for PubMed dataset. The top-performing algorithm is shown in bold, while the second top-performing algorithm is in italics.} \label{tab:real_data_2}
\begin{tabular}{lccccccc}
    \toprule
    \cmidrule(lr){1-1} \cmidrule(lr){2-7}
    Measure & LSM & CALSM & SVDy & SVDyz & CAcluster & NACcluster \\
    \midrule       
    RI & -- & \textbf{0.690} & 0.558 & 0.522 & \textit{0.681} & 0.509 \\
    Time/h & -- & 0.844 & \textit{0.172} & 0.176 & 0.661 & \textbf{0.062}  \\
    \bottomrule
\end{tabular}
   
\end{table}
\section{Discussions}\label{sec:discussion}
Several research goals are set to be explored in the future: The first problem is the heterogeneous network, where the degree of network nodes may be significantly different. For heterogeneous networks, a key question is whether posterior convergence can still achieve an improved rate. This is challenging because the number of effective parameters for node heterogeneity $n$ already exceeds $(s_x+s_b)d\log(np)$. The other research goal is to extend the proposed framework to consider multilayer networks from the same covariate. The main challenge is to determine how to exchange information between different layers of networks.

\section{Supplementary material}
The supplementary material provides algorithm details for CAVI variational inference in Section~\ref{sec:CAVI},  proofs for the main theorems in Sections~\ref{sec:proof_start} to~\ref{sec:proof_end} and additional Lemmas in Section~\ref{sec:lemmas}. Specifically, reproducible examples for both simulation and real data analysis can be found in the attached files.

% { \putbib}
% \end{bibunit}

	% \newpage
\appendix	

\renewcommand{\theequation}{A.\arabic{equation}}
\setcounter{equation}{0}

% \title{Appendix for `Robust High-Dimensional Covariate-Assisted Network Modeling'}

	\section{Appendix for `Robust High-Dimensional Covariate-Assisted
Network Modeling'}

%  \begin{bibunit}

\subsection{Derivation of the CAVI algorithm} \label{sec:CAVI}
As we adopt the variable augmentation that the square of the half-Cauchy distribution can be expressed as a mixture of Inverse-Gamma distributions \citep{neville2014mean}. 
        \begin{align*}
            \lambda_{x_i}^2 \mid v_{x_i} \sim \mbox{IG}(1/2,1/v_{x_i}),\,\, v_{x_i} \sim \mbox{IG}(1/2,1). \,\,
             \tau_{x}^2 \mid v_{x} \sim \mbox{IG}(1/2,1/v_{x}),\quad v_{x} \sim \mbox{IG}(1/2,1). \\
              \lambda_{b_j}^2 \mid v_{b_j} \sim \mbox{IG}(1/2,1/v_{b_j}),\,\,\quad v_{b_j} \sim \mbox{IG}(1/2,1).\,\,
               \tau_{b}^2 \mid v_{b} \sim \mbox{IG}(1/2,1/v_{b}),\quad v_{b} \sim \mbox{IG}(1/2,1),  
        \end{align*}
        where $\mbox{IG}(a,b)$ stands for Inverse Gamma distribution with shape parameter $a$ and scale parameter $b$.
Suppose at the step $k$, we have the following updating distribution
\begin{align*}
    q^*_{\beta_0} & = N(\mu_{\beta_0} , \sigma^2_{\beta_0} ) \\
    q^*_{x_i} & = N(\bm{\mu}_{x_i}, \Sigma_{x_i}) \\
    q^*_{b_j} & = N(\bm{\mu}_{b_j}, \Sigma_{b_j}) \\
    q^*_{\lambda^2_{x_i}} & = \text{IG}(a_{\lambda^2_{x_i}}, b_{\lambda^2_{x_i}}), \qquad
    q^*_{v_{x_i}} = \text{IG}(a_{v_{x_i}}, b_{v_{x_i}}) \\
    q^*_{\tau^2_{x}} & = \text{IG}(a_{\tau^2_{x}}, b_{\tau^2_{x}}), \qquad
    q^*_{v_{x}} = \text{IG}(a_{v_{x}}, b_{v_{x}}) \\
    q^*_{\lambda^2_{b_j}} & = \text{IG}(a_{\lambda^2_{b_j}}, b_{\lambda^2_{b_j}}), \qquad
    q^*_{v_{b_j}} = \text{IG}(a_{v_{b_j}}, b_{v_{b_j}}) \\
    q^*_{\lambda^2_{b}} & = \text{IG}(a_{\lambda^2_{b}}, b_{\lambda^2_{b}}), \qquad
    q^*_{v_{b}} = \text{IG}(a_{v_{b}}, b_{v_{b}})
\end{align*}
Then, under some derivation as in \cite{jaakkola2000bayesian}, we have the updating of the  tangent parameter
\begin{align}\label{update_xi}
\begin{aligned}
    \xi^2_{ij}  =&\text{tr}(\Sigma_{x_i}\Sigma_{x_j}) + \bm{\mu}_{x_i}^\top\Sigma_{x_j} \bm{\mu}_{x_i} + \bm{\mu}_{x_j}^\top\Sigma_{x_i} \bm{\mu}_{x_j} + (\bm{\mu}_{x_i}^\top\bm{\mu}_{x_j})^2  \\&+2\mu_{\beta_0}\bm{\mu}_{x_i}^\top\bm{\mu}_{x_j}+\mu^2_{\beta_0}+\sigma^2_{\beta_0}
    \end{aligned}
    \end{align}

    The updating of the parameter for the variational distribution of $\beta_0$
    \begin{align}\label{update_beta}
 \sigma^2_{\beta_0} = \frac{1}{1/\sigma_0^2+2\sum_{i<j}A_{ij}}, \qquad   \mu_{\beta_0}  = \frac{\sum_{i<j}(y_{ij}-\frac{1}{2}-2A_{ij}\bm{\mu}_{x_i}'\bm{\mu}_{x_j}) + \mu_0/\sigma^2_0 }{1/\sigma^2_0+2\sum_{i<j}A_{ij}}
    \end{align}

        The updating of the parameter for the variational distribution of $\+x_i$
    \begin{align}
    \begin{aligned}\label{update_X}
        \Sigma_{\+{x}_i} &= \Big\{\sum_{j\neq i}2A_{ij}\Big(\Sigma_{\+{x}_j}+\bm{\mu}_{x_j}\bm{\mu}_{x_j}^\top \Big) + \frac{a_{\lambda^2_{x_i}}a_{\tau^2_{x}}}{b_{\lambda^2_{x_i}}b_{\tau^2_{x}}}I_d  \Big\}^{-1} \\
    \bm{\mu}_{x_i} & = \Sigma_{x_i}\Big\{ \sum_{j\neq i} \Big(y_{ij}-\frac{1}{2}-2A_{ij}\mu_{\beta_0} \Big)\bm{\mu}_{x_j} + \frac{a_{\lambda^2_{x_i}}a_{\tau^2_{x}}}{b_{\lambda^2_{x_i}}b_{\tau^2_{x}}} (\bm{\mu}_{b_1},\ldots,\bm{\mu}_{b_p}) \+{z}_i \Big\}
    \end{aligned}
    \end{align}

    The updating of the parameter for the variational distribution of scales of node $i$
    \begin{align}
  &  a_{\lambda_{x_i}}  = \frac{d+1}{2}, \qquad   b_{\lambda_{x_i}}  =  \frac{1}{2}\frac{a_{\tau^2_x}}{b_{\tau^2_x}}\Big\{\text{tr}(\Sigma_{x_i})+\bm{\mu}_{x_i}^\top\bm{\mu}_{x_i} 
+ \+{z}_i^\top\Big[\text{diag}\big\{\text{tr}(\Sigma_{b_1}),\ldots,\text{tr}(\Sigma_{b_p})\big\} \nonumber \\
&+ (\bm{\mu}_{b_i}^\top\bm{\mu}_{b_j})_{ij}\Big]\+{z}_i-2\+{z}_i^\top(\bm{\mu}_{b_1},\ldots,\bm{\mu}_{b_p})^\top\bm{\mu}_{x_i} \Big\} + \frac{a_{v_{x_i}}}{b_{v_{x_i}}}  \nonumber \\
  &  a_{v_{x_i}}  = 1, \qquad   b_{v_{x_i}} = 1 + \frac{a_{\lambda^2_{x_i}}}{b_{\lambda^2_{x_i}}} \label{update_scale_X}\\
   & a_{\tau_{x}}  = \frac{nd+1}{2}, b_{\tau_{x}}  = \sum_{i=1}^n\frac{1}{2}\frac{a_{\lambda^2_{x_i}}}{b_{\lambda^2_{x_i}}}\Big\{\text{tr}(\Sigma_{x_i})+\bm{\mu}_{x_i}^\top\bm{\mu}_{x_i} 
+ \+{z}_i^\top\Big[\text{diag}\big\{\text{tr}(\Sigma_{b_1}),\ldots,\text{tr}(\Sigma_{b_p})\big\} \nonumber\\
&+ (\bm{\mu}_{b_i}^\top\bm{\mu}_{b_j})_{ij}\Big]\+{z}_i-2\+{z}_i^\top(\bm{\mu}_{b_1},\ldots,\bm{\mu}_{b_p})^\top\bm{\mu}_{x_i} \Big\} + \frac{a_{v_{x}}}{b_{v_{x}}}  \nonumber\\
  &  a_{v_{x}}  = 1, \qquad   b_{v_{x}}  = 1 + \frac{a_{\tau^2_{x}}}{b_{\tau^2_x}} \nonumber
    \end{align}

        The updating of the parameter for the variational distribution of $\+b_j$
    \begin{align}\label{update_B}
        \begin{aligned}
            \Sigma_{\+{b}_j} & = \Bigg\{ \Big(\sum_{i=1}^n\frac{a_{\lambda^2_{x_i}}a_{\tau^2_{x}}}{b_{\lambda^2_{x_i}}b_{\tau^2_{x}}}z^2_{ik} + \frac{a_{\lambda^2_{b_j}}a_{\tau^2_b}}{b_{\lambda^2_{b_j}}b_{\tau^2_b}}\Big)I_d \Bigg\}^{-1}  \\
    \bm{\mu}_{\+{b}_j} & = \Sigma_{\+{b}_j} \Bigg\{ \sum_{i=1}^n z_{ik}\frac{a_{\lambda^2_{x_i}}a_{\tau^2_x}}{b_{\lambda^2_{x_i}}b_{\tau^2_x}} \Big( \sum_{l\neq k}z_{il}\bm{\mu}_{b_l} - \bm{\mu}_{x_i}\Big)\Bigg\}
   \end{aligned}
    \end{align}

        The updating of the parameter for the variational distribution of scales of covariate $k$
    \begin{align}
    \begin{aligned}\label{update_scale_B}
    a_{\lambda_{b_j}} & = \frac{d+1}{2}, \qquad
    b_{\lambda_{b_j}}  = \frac{1}{2}\frac{a_{\tau^2_b}}{b_{\tau^2_b}}\Big\{ \text{tr}(\Sigma_{b_j}) + \bm{\mu}_{b_j}^\top\bm{\mu}_{b_j} \Big\} + \frac{a_{v_{b_j}}}{b_{v_{b_j}}}  \\
    a_{v_{b_j}} & = 1, \qquad
    b_{v_{b_j}} =  1 + \frac{a_{\lambda^2_{b_j}}}{b_{\lambda^2_{b_j}}} \\
    a_{\tau_{b}} & = \frac{pd+1}{2}, \qquad
    b_{\tau_{b}}  = \sum_{k=1}^p\frac{1}{2}\frac{a_{\lambda^2_{b_j}}}{b_{\lambda^2_{b_j}}^2}\Big\{ \text{tr}(\Sigma_{b_j}) + \bm{\mu}_{b_j}^\top\bm{\mu}_{b_j} \Big\} + \frac{a_{v_b}}{b_{v_b}} \\
    a_{v_{b}} & = 1, \qquad
    b_{v_{b}}  = 1 + \frac{a_{\tau^2_b}}{b_{\tau^2_b}}
    \end{aligned}
\end{align}

% \subsection{Computation time of large scale networks}\label{sec:time}

% In this subsection, we present the computation time of the proposed algorithm for networks with nodes of the order of $O(10^4)$. In particular, we let $n=10000$, change the number of covariates $p$ and show the corresponding time spent in hours. We used the same settings as Case 2 in section~\ref{sec:data_analysis} (only change $n$ and $p$) and the same CAVI algorithm setting in our approach, with a stopping criterion of either 500 cycles or a mean difference of less than $1e-4$ between the estimated denoised connecting probabilities in two consecutive cycles. All reported results were obtained on a Windows 11 PC with an Intel(R)
% Core(TM) Processor (i9-14900 v3@2.00GHz) and 32G RAM.

% \begin{table}[]
%     \centering
%     \begin{tabular}{c|ccccc}
%     \hline
%        $p$  &   100 &200 &500 &1000 &2000\\
%         \hline
%        Time in hours  & 7.47 & 8.02 & 8.29
% & 10.52 & 20.01\\
%         \hline
%     \end{tabular}
%     \caption{ Time cost in hours for networks with a number of nodes $n=10000$ and the varying number of covariates $p$.}
%     \label{tab:my_label}
% \end{table}

\subsection{Proof of Theorem~\ref{thm:posterior}}\label{sec:proof_start}

\textbf{Upper bound of KL ball:} As shown in Theorem 3.2 in \cite{bhattacharya2019bayesian}, under the condition that 
	\begin{equation*}
	\Pi(B_n(\varTheta^*;\epsilon_{n})) \geq e^{-n^2\epsilon_{n}^2}, 
	\end{equation*}
	we can obtain the convergence of the $\alpha$-divergence at the rate:
	\begin{equation*}
	D_\alpha(p_{\varTheta},p_{\varTheta^*}) = \frac{1}{\alpha-1} \log \int (p_{\varTheta^*})^\alpha (p_{\varTheta})^{1-\alpha} d\mu.
	\end{equation*}

	Denote the $\epsilon$ ball for KL divergence neighborhood centered at $\varTheta^*$ as
	\begin{equation*}
	B_n(\varTheta^*;\epsilon) = \left\{\varTheta\in \Xi: \int p_{\varTheta^*}\log(\frac{p_{\varTheta^*}}{p_{\varTheta}}) d\mu \leq n^2 \epsilon^2, \int p_{\varTheta^*}\log^2(\frac{p_{\varTheta^*}}{p_{\varTheta}}) d\mu \leq n^2 \epsilon^2\right\},
	\end{equation*}
	where $\mu$ is the Lebesgue measure and 
 \begin{equation*}
     D_{KL}(p_{\varTheta},p_{\varTheta^*})=\int p_{\varTheta^*}\log(\frac{p_{\varTheta^*}}{p_{\varTheta}}) d\mu,  \quad V_2(p_{\varTheta},p_{\varTheta^*}) =  \int p_{\varTheta^*}\log^2(\frac{p_{\varTheta^*}}{p_{\varTheta}}) d\mu.
 \end{equation*}
	
	First, by  Assumption~\ref{asm:sparsity_prior}, we have for the sparsity parameter $\beta$ (an example is provided in Lemma~\ref{lem:scale}), we have 
\begin{align*}
  \Pi ( |\beta-\beta^*|^2 \leq \epsilon_n^2) 
  \gtrsim \exp \left(-K'n^2\epsilon_n^2\right),
\end{align*}
for some constant $K'>0$.

Denote $E_0 := \{\beta: |\beta-\beta^*|^2 \leq \epsilon_n^2\}$. Conditional on $E_0$, based on Lemma~\ref{lem:binary_KL} and~\ref{lem:second_KL}, we have $$\max\{D_{KL}(p_{\varTheta},p_{\varTheta^*}),V_2(p_{\varTheta},p_{\varTheta^*})\}\lesssim  \|\+X\+X'-\+X^* \+X^{*'}+\beta \+1 \+1' -\beta^* \+1\+1'\|_F^2.$$
 
 Hence after conditioning on $E_0$, we only need to lower bound the prior probability of the set  $E_1:=\{\sum_{i,j}(\+x_{i}'\+x_{j}-\+x^{*'}_{i}\+x^{*}_{j})^2\leq n^2 \epsilon^2_n\} $. Given $i,j$ we have
	\begin{equation*}
	\begin{aligned}
	|\+x_{i}'\+x_{j}-\+x^{*'}_{i}\+x^{*}_{j}| &\leq |(\+x'_{i}-\+x^{*'}_{i})\+x^{*}_{j}|+|\+x'_{i}(\+x_{j}-\+x^{*}_{j})|= |(\+x'_{i}-\+x^{*'}_{i})\+x^{*}_{j}|+|(\+x'_{i}-\+x^{*'}_{i}+\+x^{*'}_{i})(\+x_{j}-\+x^{*}_{j})| \\
	&\leq  \|\+x_{i}-\+x^{*}_{i}\|_2\|\+x^*_{j}\|_2 + \|\+x_{j}-\+x^{*}_{j}\|_2\|\+x^*_{i}\|_2 + \|\+x_{i}-\+x^{*}_{i}\|_2\| \|\+x_{j}-\+x^{*}_{j}\|_2
	\end{aligned}
	\end{equation*}
 Since $\|\+x^*_{i}\|_2 \lesssim 1$ for all $i$ by Assumption~\ref{asm:bound}, together with $n\epsilon_n^2 \lesssim 1$ under the condition $s_b \lesssim n$.  We have
 % Condition on the event $\{\max_i \|\+x_i-\+x_i^*\|_2 \leq \epsilon\}$,
 % We have 
 % $$ |\+x_{i}'\+x_{j}-\+x^{*'}_{i}\+x^{*}_{j}| \lesssim  \epsilon,$$ which implies 
 $$ \{\sum_{i,j}(\+x_{i}'\+x_{j}-\+x^{*'}_{i}\+x^{*}_{j})^2 \lesssim n^2\epsilon^2_n\} \supseteq \{  \sum_{i \in [n]} \|\+x_i-\+x_i^*\|_2^2 \leq n\epsilon^2_n\} 	 $$ Hence, it suffices to consider the event
 \begin{equation*}
     E_{10} : = \{\+x_i: \sum_{i \in [n]} \|\+x_i-\+x_i^*\|_2^2  \leq  n\epsilon^2_n= \max\{1,s_b+s_x\} d\log(np)/n\}.
 \end{equation*}
Denote $E_{11} : =\{\sum_{i \in S} \|\+x_i - \+x_i^{*'}\|_2^2 \leq \max\{1,s_x\}d\log(n)/(2n)\}$ and $E_{12} : =\{\sum_{i \in S^c} \|\+x_i - \+x_i^{*'}\|_2^2 \leq  \max\{1,s_x+s_b\}d\log(np)/(2n) \}$. Clearly, $E_{10} \supset E_{11} \cap E_{12} $. Denote  $\tilde {\+x}_i^* = \+B^{*'} \+z_i$.  Let $$\delta = \sqrt{\frac{s_x d\log n}{2n}}.$$

 		Denote \begin{equation}
		\tau_x^\ast = \frac{s_x^{\frac{1}{2}}(\log (n))^{\frac{1}{2}}}{n^{2}d^{1/2}},
		\end{equation}
		and consider the event $E_{\tau_x} = \{\tau_x:\tau_x \in [\tau_x^\ast,2\tau_x^\ast]\}$.

\textbf{Concentration for index $S$:} 
Consider the set $S$ such that $\+x_i^* \ne \tilde {\+x}_i^*$ for $i \in S$. Let $\delta_0 = \delta/\sqrt{s_x}$. The joint density of $x_i$ and $\lambda_i$ given $\tau_x$ is:

$$
p\left(\+x_i, \lambda_{x_i} \mid \tau_x\right)=\frac{2}{\pi \times (2 \pi)^{d / 2}} \exp \left(-\frac{\left\|\+x_i\right\|_2^2}{2 \lambda_{x_i}^2 \tau_x^2}\right) \frac{1}{\lambda_i^d \tau_x^d\left(1+\lambda_i^2\right)}.
$$
Then we have 
			\begin{equation*}
		\begin{aligned}
	&	Pr(\|\+x_i-\+x_i^*\|_2 < \delta_0 \mid \tau_x) =\int_{\lambda_i} \int_{\|\+x_i-\+x_i^*\|_2<\delta_0} 2^{1-d/2} \pi^{-1-d/2}\exp\left\{-\frac{\|\+x_i\|_2^2}{2\lambda_{x_i}^2 \tau_x^2}\right\} \frac{1}{\lambda_i^d \tau_x^d (1+\lambda_i^2)}  d\lambda_{x_i} d \+x_i \\
		&\geq 2^{1-d/2} \pi^{-1-d/2}\int_{\|\+x_i-\+x_i^*\|_2<\delta_0}\int_{1 /\tau^\ast}^{2 /\tau^\ast} \exp\left\{-\frac{\|\+x_i\|_2^2}{2\lambda_{x_i}^2 \tau_x^2}\right\} \frac{1}{\lambda_i^d \tau_x^d (1+\lambda_i^2)}  d\lambda_{x_i} d \+x_i\\
		&\stackrel{(i)}{\geq} 2^{1-d/2} \pi^{-1-d/2}   \int_{\|\+x_i-\+x_i^*\|_2<\delta_0}\exp\left\{-\frac{\|\+x_i\|_2^2}{2 }\right\} \int_{1 /\tau^*}^{2 /\tau^*}  \frac{1}{4^d   ( 1+\lambda_i^2)}d\lambda_{x_i} d  \+x_i \\
		& \geq 2^{1-d/2} \pi^{-1-d/2} 4^{-d} \frac{\tau^{*2}}{\tau^{*2}+4}  \int_{\|\+x_i-\+x_i^*\|_2<\delta_0} \exp\left\{-\frac{\|\+x_i\|_2^2}{2}\right\}d \+x_i\\
		& \stackrel{(ii)}{\geq} K \exp(-cd) \tau_x^{*2}  \delta_0^d \\
		& {\geq } K'\exp(-cd) \frac{s_x \log p}{n} \frac{(\log n)^{d/2}}{n^d} \geq K'' (n)^{-Md}
		\end{aligned}
		\end{equation*}
		where $(i)$ holds based on $1 <\tau_x \lambda_{x_i}<4  $ for the range of $\tau$ and $\lambda_i$; $(ii)$ is due to $\tau<1$ and $\max_j (\|\+x_i\|_2) = O(\sqrt{d})$; $(iii)$ is because $s_x<n$; $M,K,K',K''>0$ are some constants.
  Then by considering all the events for $i \in S$, we get
\begin{equation} \label{eq:assist_1}
    \Pi(\sum_{i \in S} \|\+x_i - \+x_i^{*'}\|_2^2 \leq C\delta^2 \mid E_{\tau_x}) \gtrsim e^{-M s_x d \log n},
\end{equation}
for large enough constant $C>0$.

\textbf{Concentration for index $S^c$:} 
In addition, for the index $S^c$ such that $\+x_i^* = \+B^{*'} \+z_i$ for $i \in S^c$. Then conditional on the event $E_{20}$, we have
 Note that 
  \begin{align*}
      \|\+x_{i}-\+x^{*}_{i}\|_2 \leq  \|\+x_{i}-\+B^{*'} \+z_i \|_2+ \| \+B^{*'}\+z_i-\+x^{*}_{i}\|_2  =  \|\+x_{i}-\+B^{*'} \+z_i \|_2 \\
\leq    \|\+x_i-\+B\+z_i\|_2   + \|(\+B^*-\+B)'\+z_i \|_2.      
\end{align*}
By Lemma~\ref{lem:horse_net}, denote $$E_{20} :=\{\|\+B-\+B^*\|_{2,2} \leq \sqrt{\max\{s_b,1\} d\log p/(n^2)}\},$$ then we have $\log \Pi(E_{20}) \gtrsim -\max\{s_b,1\} d\log(np)$. Then conditional on the event $ E_{20}$, we have 
\begin{equation*}
    \Pi( E_{12} \mid E_{\tau_x}) \geq  \Pi(E_{12} \mid E_{20}, E_{\tau_x}) \Pi(E_{20}).
\end{equation*}
Conditional on $E_{20}$ and $E_{\tau_x}$,  it is sufficient to calculate the concentration of the following prior
\begin{equation*}
N(\+B\+z_i,\tau_x^2\lambda_{x_i}^2) \mid \lambda_{x_i} \sim \mbox{Ca}^+(0,1)
\end{equation*}
around $\+B\+z_i$.

		By the Chernoff type of bound for $d$-dimension Gaussian random vectors (e.g., Lemma 1 in \cite{jin2019short}), we have, we have
		\begin{equation*}
		\Pi(\|\+x_i-\+B\+z_i\|_2> \frac{\delta}{\sqrt{n}} \mid \lambda_{x_i},\tau_x) \leq  2e^{-\delta^2/(2nd\lambda_{x_i}^2\tau_x^2)}.
		\end{equation*}
		Then we have
		\begin{equation*}
		\begin{aligned}
		\Pi(\|\+x_i-\+B\+z_i\|_2< \frac{\delta}{2\sqrt{n}} \mid \tau_x) &\geq \int_{\lambda_{x_i}} \left\{1-2e^{-\delta^2/(2nd \lambda_{x_i}^2\tau_x^2)}\right\} f(\lambda_{x_i} ) d\lambda_{x_i} \\
		&= 1- \int_{\lambda_{x_i}} 2e^{-\delta^2/(2nd \lambda_{x_i}^2\tau_x^2)} f(\lambda_{x_i} ) d\lambda_{x_i},
		\end{aligned}
		\end{equation*}
		with $f(\lambda_{x_i} ) = 1/(1+\lambda_{x_i}^2) <\lambda_{x_i}^{-2}$. Then, we have
		\begin{equation*}
		\begin{aligned}
		\int_{\lambda_{x_i}} 2e^{-\delta^2/(2nd \lambda_{x_i}^2\tau_x^2)} f(\lambda_{x_i}) d\lambda_{x_i} &<  2 \int_{\lambda_{x_i}} e^{-\delta^2/(2nd \lambda_{x_j}^2\tau_x^2)} \lambda_{x_i}^{-2}  d\lambda_{x_i} \\
		& = 2 \frac{\Gamma(1/2)}{\{\delta^2/(2nd \tau_x^2)\}^{1/2}} \\
		& = C \frac{\sqrt{nd}\tau_x}{\delta} =C \frac{n\tau}{\sqrt{s_x\log n} } \\
		& {\leq} C' \frac{1}{n},
		\end{aligned}
		\end{equation*}
		where in the final step we use $\tau_x < 2\tau_x^*$.

Therefore, conditional on $E_{20}$,  by product all the probability in $i \in S^c$, we have 
\begin{equation*}
  \Pi(E_{12} \mid E_{20}, E_{\tau_x})  \gtrsim (1-\frac{1}{n})^n \gtrsim  e^{-\max\{1,s_x\}\log n}.
\end{equation*}
Note that
\begin{align*}
\begin{aligned}
 \sum_{i \in S^c} \|\+x_i-\+B^*\+z_i\|^2_2 &\lesssim  \sum_{i \in S^c} \|\+x_i-\+B\+z_i\|^2_2   + \|(\+B^*-\+B)'\+Z \|_F^2 
 \end{aligned}
\end{align*}
Therefore by Assumption~\ref{asm:restricted_eig},
\begin{align} \label{eq:assist_2}
\begin{aligned}
     \Pi(E_{12} \mid  E_{\tau_x})  \geq    \Pi(E_{12} \mid E_{20}, E_{\tau_x})  \times  \Pi( E_{20})  \\ \gtrsim \exp(-C\max\{1,s_x+s_b\}d \log(np)), 
    \end{aligned}
\end{align}
for some constant $C>0$.

Note that \begin{equation}\label{eq:assist_3}
   \Pi(E_{\tau_x}) = \int_{\tau^*}^{2\tau_x^*}\frac{2}{\pi} \frac{1}{1+\tau^2} \geq \tau_x^* \times \frac{2}{\pi} \frac{1}{1+4\tau_x^{*2}} \geq C \tau_x^* \gtrsim e^{-c \log n} 
\end{equation} for constants $C,c>0$.

\textbf{Aggregating all results:} 
Then by aggregating the results of equation~\eqref{eq:assist_1}, equation~\eqref{eq:assist_2} and equation~\eqref{eq:assist_3}, we have

\begin{align*}
  &\Pi ( \|\+X-\+X^*\|^2_F \leq n\epsilon_n^2) = \Pi(E_{11} \mid E_{\tau_x}) \times \Pi(E_{12} \mid E_{20}, E_{\tau_x}) \times  \Pi(E_{20})\times  \Pi(E_{\tau_x}) \\
  &\geq \left\{\Pi \left(\sum_{i \in S} \|\+x_i-\+x^*_i\|^2_2 \leq \max\{1,s_x\}d\log(n)/(2n) \mid E_{\tau_x} \right)\right. \\
  &\times \Pi \left.\left( \sum_{i \in S^c} \|\+x_i-\+B^*\+z_i\|^2_2 \leq \max\{1,s_x+s_b\}d\log(np)/(2n) \mid E_{\tau_x}\right)\right\} \Pi(E_{\tau_x}) \\
  &\gtrsim \exp \left(-Kn^2\epsilon_n^2\right),
\end{align*}
for large enough constant $K>0$.

Therefore, the final conclusion follows from the above prior mass results and Theorem 3.2 in \cite{bhattacharya2019bayesian}.

\subsection{Proof of Corollary~\ref{thm:posterior_latent_vector}}
As Theorem~\ref{thm:posterior} works for any $\alpha \in (0,1)$, it works for $\alpha=1/2$. Based on Lemma~\ref{lem:lower of divergence} in the appendix, we have for $a,b$ such that $|a|,|b| \gtrsim 1$ and $a,b<0$,  we have
 		\begin{align*}
		D_{\frac{1}{2}}(P_a,P_b)\gtrsim (\sqrt{p_a}-\sqrt{p_b})^2,
		\end{align*}
  and
\begin{equation*}
		D_{\frac{1}{2}}(P_a,P_b) \gtrsim  \exp \left\{a \wedge b \right\}(b-a)^2.
\end{equation*}

Denote $p^*_{ij} = 1/\{1+\exp(-\beta^*-\+x_{i}^{*'}\+x_{j}^{*})\}$, $\hat{p}_{ij} = 1/\{1+\exp(-\hat{\beta}-\hat {\+x}_{i}^{'}\hat {\+x}_{j})\}$, $p^*_{\beta} = 1/\{1+\exp(-\beta^*)\}$ and $\hat p_{\hat \beta} = 1/\{1+\exp(-\hat \beta)\}$. Due to the uniform boundedness of $\+x_{i}^*$ and $\hat{\+x}_{i}$, we have there are uniform constants $c,C$ such that
\begin{equation*}
c p^*_{\beta}   \leq p^*_{ij} \leq Cp^*_{\beta}; \quad c \hat p_{\hat \beta} \leq \hat{p}_{ij}  \leq C  \hat p_{\hat \beta}.
\end{equation*}
Then we have 
\begin{equation*}
\begin{aligned}
    \frac{1}{n^2}D_{\frac{1}{2}}(P_{\m{\hat{X}}, \hat \beta},P_{\m X^*, \beta^*}) = \frac{1}{n^2}\sum_{i,j=1}^n D_{\frac{1}{2}}(p^*_{ij},\hat{p}_{ij})\\
    \gtrsim  \frac{1}{n^2}\sum_{i,j=1}^n (\sqrt{p^*_{ij}}-\sqrt{\hat{p}_{ij}})^2\\
    \gtrsim (C \sqrt{p^*_{\beta} \vee \hat{p}_{ \hat \beta}}-c\sqrt{p^*_{\beta} \wedge \hat{p}_{\hat \beta}} )^2.
    \end{aligned}
\end{equation*}
If $\hat{\beta} \ll \beta^* $, which means $\hat p_{\hat{\beta}} \ll p_{\beta^*}$, then we have $\epsilon_{n}^2 \gtrsim D_{\frac{1}{2}}(P_{ \hat{\m X}, \hat \beta},P_{\m X^*, \beta^*})/ (n^2) \gtrsim (c\sqrt{\hat p_{\hat{\beta}}}-C\sqrt{p_{\beta^*}})^2 \gtrsim p_{\beta^*}$, which causes contradiction with Assumption~\ref{asm:sparsity_level}. Therefore, we must have $\hat{\beta} \gtrsim \beta^*$, then
 \begin{align*}
     \epsilon_{n}^2 \gtrsim \frac{1}{n^2}D_{\frac{1}{2}}(P_{\m{\hat{X}}, \hat \beta},P_{\m X^*, \beta^*})\\
     \gtrsim \frac{1}{n^2}\sum_{i,j=1}^n \exp \left\{\hat \beta \wedge \beta^* \right\}(\hat \beta-\beta^*+\hat {\+x}_{i}' \hat {\+x}_{j}-\+x_{i}^{*'}\+x_{j}^{*})^2 \\
     \gtrsim \frac{1}{n^2}\exp(\beta^*) \sum_{i,j=1}^n (\hat \beta-\beta^*+\hat {\+x}_{i}' \hat{\+x}_{j}-\+x_{i}^{*'}\+x_{j}^{*})^2 .
 \end{align*}
Finally, 
\begin{equation*}
\begin{aligned}
    &\sum_{i, j=1}^n (\hat \beta-\beta^*+\hat {\+x}_{i}' \hat{\+x}_{j}-\+x_{i}^{*'}\+x_{j}^{*})^2 \\
    = &\sum_{i,j=1}^n (\hat {\+x}_{i}' \hat{\+x}_{j}-\+x_{i}^{*'}\+x_{j}^{*})^2 + 2\sum_{i,j=1}^n \langle \hat \beta-\beta^*, \hat {\+x}_{i}' \hat{\+x}_{j}-\+x_{i}^{*'}\+x_{j}^{*} \rangle + \sum_{i,j=1}^n(\hat \beta-\beta^*)^2.
    \end{aligned}
\end{equation*}
Notice that $\sum_{i,j} \+x_{i}^{*'}\+x_{j}^{*} = \sum_i \+x_{i}^{*'} \sum_j \+x_{j}^{*} = \+0_d$. In addition, by the Assumption~\ref{asm:projection}, $|\sum_{i,j=1}^n \hat {\+x}_{i}' \hat{\+x}_{j}| \leq  |\sum_i \hat{\+x}_{i}^{'} \sum_j \hat{\+x}_{j}|  \lesssim 1 \lesssim n\epsilon_n e^{-\beta^*/2}$. Therefore, by Young's inequality, we have

\begin{equation*}
\begin{aligned}
    &\sum_{i,j=1}^n (\hat \beta-\beta^*+\hat {\+x}_{i}' \hat{\+x}_{j}-\+x_{i}^{*'}\+x_{j}^{*})^2 
    \\ \geq &\sum_{i,j=1}^n (\hat {\+x}_{i}' \hat{\+x}_{j}-\+x_{i}^{*'}\+x_{j}^{*})^2 +(1-\frac{1}{C})   \sum_{i,j=1}^n(\hat \beta-\beta^*)^2 - C n^2\epsilon_n^2 e^{-\beta^*},
    \end{aligned}
\end{equation*}
for large enough constant $C>0$. Hence, the final conclusion holds.

\subsection{Proof of Theorem~\ref{thm:single_cluster}}\label{sec:proof_end}

	Denote 
	\begin{equation*}
	\hat{\+O} = \argmin_{\+O \in \mathbb{O}^{d \times d}} \|\hat{\+X}- \+X^* \+O \|_F.
	\end{equation*}
	Note that $\+X^* \+O$ has the same cluster configuration as $\+X^*$. By Assumption 2 in Theorem~\ref{thm:single_cluster}, we have $\sigma_d(\+X) \gtrsim \sqrt{n}$. 
	Therefore, by Lemma~\ref{lem:perturbation}, we have:
	
	\begin{equation*}
	\|\hat{\+X}- \+X^* \+O \|_F \lesssim   \|\hat{\+X} \+X'- \+X^* \+X^{*'}\|_F/\sqrt{n} \lesssim e^{-\beta^*/2}\epsilon_n.
	\end{equation*}
	Note that the smallest cluster size is greater than $16 \|\hat{\+X}- \+X^* \+O \|_F^2 /\delta^2$.	By Lemma~\ref{lem:k_means}, the above bounds implies that by applying $K$ means on $\hat{\+X}$ to obtain $\hat{\+\Xi}$, we have
	\begin{equation*}
	L(\hat{\+\Xi},\+\Xi) \leq 16 \|\hat{\+X}- \+X^* \+O \|_F^2 /\delta^2.
	\end{equation*}
	The above two inequalities lead to 
	\begin{equation*}
	 L(\hat{\+\Xi},\+\Xi) \lesssim  \frac{ \|\hat{\+X} \hat{\+X}'- \+X^* \+X^{*'}\|^2_F}{n  \delta ^2}  \lesssim \frac{e^{-\beta^*}\epsilon_n^2}{  \delta^2}.
	\end{equation*}
	Therefore, the final conclusion holds.
\subsection{Additional lemmas}\label{sec:lemmas}

Consider the prior concentration for the following shrinkage prior:  
	\begin{align}\label{shrinkage}
\+b_{i} &\sim \mathcal{N}(\+0, \lambda_{b_i}^2\tau_{b}^2 \+I_d),
        &\tau_{b} \sim \mbox{Ca}^+(0,1), \lambda_{b_i} \sim \mbox{Ca}^+(0,1), \quad i \in [p],
	\end{align}
	In the following Lemma, we prove the $\ell_{2,1}$ concentration of  the proposed prior for group sparsity truth.
	\begin{lemma}[$\ell_{2,1}$ concentration for shrinkage prior]\label{lem:horse_net}
		Suppose $\+B^* \in \mathbb{R}^{p \times d}$ with $S=\{j: \+b_j^*\ne \+0\}$ and $|S| \leq s_b$, $1\leq s_b\leq p$. Denote $\delta=\sqrt{s_b d\log (p)/(n^2)}$ with $s_b d\log p =o (n^2)$,  $\max_{j=1}^p \|\+b_j^*\|_2 =O(p^\alpha )$ for some constant $\alpha>0$. Under the above shrinkage prior~\eqref{shrinkage},
		then for some constant $c_0>0$, we have
		\begin{equation}\label{eq:hs_net}
		\Pi(\|\+B-\+B^*\|_{2,1} \leq \delta) \gtrsim e^{-K s_b d\log (np)},
		\end{equation}
		where $K>0$ is a constant.
	\end{lemma}

	\begin{proof}
		$ s_b d\log p =o (n^2)$ guarantees $\delta =o(1)$.

		First,  we have
		\begin{equation*}
		Pr(\|\+B-\+B^*\|_{2,1} <\delta) \geq \prod_{j \in S} Pr(\|\+b_j-\+b_j^*\|_2< \frac{\delta}{2s})\prod_{j \in S^c} Pr(\|\+b_j-\+b_j^*\|_2< \frac{\delta}{2p}),
		\end{equation*}
		
		Denote \begin{equation}
		\tau_b^\ast = \frac{s_b^{\frac{1}{2}}(\log (p))^{\frac{1}{2}}}{n p^2d^{\frac{1}{2}}},
		\end{equation}
		and consider the event $E_{\tau_b} = \{\tau_b:\tau_b \in [\tau_b^\ast,2\tau_b^\ast]\}$.
		
		For the non-signal part. For $j \in S^c$, by the Chernoff type of bound for $d$-dimension Gaussian random vectors (e.g., Lemma 1 in \cite{jin2019short}), we have
		\begin{equation*}
		Pr(\|\+b_j \|_2> \frac{\delta}{2p} \mid \lambda_{b_j},\tau) \leq  2e^{-\frac{\delta^2}{8p^2\lambda_{b_j}^2\tau_b^2 d}}.
		\end{equation*}
		Then we have
		\begin{equation*}
		\begin{aligned}
		Pr(\|\+b_j \|_2< \frac{\delta}{2p} \mid \tau_b) &\geq \int_{\lambda_j} \left\{1-2e^{-\frac{\delta^2}{8p^2\lambda_{b_j}^2\tau_b^2 d}}\right\} f(\lambda_j ) d\lambda_{b_j} \\
		&= 1- \int_{\lambda_j} 2e^{-\frac{\delta^2}{8p^2\lambda_{b_j}^2\tau_b^2 d}} f(\lambda_j ) d\lambda_{b_j},
		\end{aligned}
		\end{equation*}
		with $f(\lambda_j ) = 1/(1+\lambda_j^2) <\lambda_j^{-2}$. Then, we have
		\begin{equation*}
		\begin{aligned}
		\int_{\lambda_j} 2e^{-\frac{\delta^2}{8p^2\lambda_{b_j}^2\tau_b^2 d}}f(\lambda_j) d\lambda_{b_j} &<  2 \int_{\lambda_j} e^{-\frac{\delta^2}{8p^2\lambda_{b_j}^2\tau_b^2 d}}\lambda_j^{-2}  d\lambda_{b_j} 
  & = 2 \int_{t=0}^\infty e^{-\frac{\delta^2 t}{8p^2\tau_b^2 d}}t^{-1/2}  dt\\
		& = 2 \frac{\Gamma(1/2)}{\{\delta^2/(8p^2\tau_b^2 d)\}^{1/2}} \\
		& = C \frac{p\tau_b \sqrt{d}}{\sqrt{s_b\log (p)/(n^2)}}\\
		& {\leq} C' \frac{1}{p},
		\end{aligned}
		\end{equation*}
		where in the final step we use $\tau < 2\tau_b^*$.
		
		Moreover, for the signal part, let $\delta_0 =\delta/(2s)$, we have
		\begin{equation*}
		\begin{aligned}
	&	Pr(\|\+b_j-\+b_j^*\|_2 < \delta_0 \mid \tau_b) =\int_{\lambda_j} \int_{\|\+b_j-\+b_j^*\|_2<\delta_0} 2^{1-d/2} \pi^{-1-d/2}\exp\left\{-\frac{\|\+b_j\|_2^2}{2\lambda_{b_j}^2 \tau_b^2}\right\} \frac{1}{\lambda_j^d \tau_b^d (1+\lambda_j^2)}  d\lambda_{b_j} d \+b_j \\
		&\geq 2^{1-d/2} \pi^{-1-d/2}\int_{\|\+b_j-\+b_j^*\|_2<\delta_0}\int_{p^\alpha /\tau^\ast}^{2 p^\alpha /\tau^\ast} \exp\left\{-\frac{\|\+b_j\|_2^2}{2\lambda_{b_j}^2 \tau_b^2}\right\} \frac{1}{\lambda_j^d \tau_b^d (1+\lambda_j^2)}  d\lambda_{b_j} d \+b_j\\
		&\stackrel{(i)}{\geq} 2^{1-d/2} \pi^{-1-d/2}   \int_{\|\+b_j-\+b_j^*\|_2<\delta_0}\exp\left\{-\frac{\|\+b_j\|_2^2}{4p^{2\alpha} }\right\} \int_{p^\alpha /\tau^*}^{2p^\alpha /\tau^*}  \frac{1}{4 p^{d\alpha}  ( 1+\lambda_j^2)}d\lambda_{b_j} d  \+b_j \\
		& \geq 2^{1-d/2} \pi^{-1-d/2} \frac{\tau^*}{4p^{(d-1)\alpha}(4p^{2\alpha}+\tau^{*2})} \int_{\|\+b_j-\+b_j^*\|_2<\delta_0} \exp\left\{-\frac{\|\+b_j\|_2^2}{4p^{2\alpha}}\right\}d \+b_j\\
		& \stackrel{(ii)}{\geq} K \delta_0^d \tau_b^* p^{-(d+1)\alpha} \\
		& {\geq } K' \left(\frac{d\log (p)}{n^2s}\right)^{d/2} p^{-(d+1)\alpha} \frac{s_b^{\frac{1}{2}}(\log (p))^{\frac{1}{2}}}{np^2} \stackrel{(iii)}{\geq} K'' (np)^{-Md}
		\end{aligned}
		\end{equation*}
		where $(i)$ holds based on $p^\alpha <\tau \lambda_{b_j}<4 p^\alpha $ for the range of $\tau$; $(ii)$ is due to $\tau<1$ and $\max_j (\|\+b_j\|_2) = O(p^\alpha )$; $(iii)$ is because $s_b<p$; $M,K,K',K''>0$ are some constants.
		
		%		In addition,  
		
		Therefore, we have
		\begin{align*}
		\Pi(\|\+B-\+B^*\|_{2,1} <\delta ) \geq	\Pi(\|\+B-\+B^*\|_{2,1} <\delta \mid E_{\tau_b})\Pi(E_{\tau_b}) \\
  \geq (1-C'/(p))^{p-s_b} K'' e^{-M s_b d\log (np)}   \gtrsim e^{-K^* \max\{1,s_b\} d\log (np)},
		\end{align*}
		where $K^*$ is a positive constant.
		
	\end{proof}

 \begin{lemma}[Prior of the scales]\label{lem:scale}
     Suppose we use the prior $\m N(0,\sigma^2_\beta)$ prior for $\beta$ with $\sigma_\beta= \sqrt{\log n}$. If Assumption~\ref{asm:sparsity_level} is satisfied, then we have $\log(\Pi(|\beta-\beta^*|\leq \epsilon_n  )) \gtrsim  -n^2\epsilon_{n}^2$ for any $0 \leq s_x,s_b \lesssim n$.
 \end{lemma}
\begin{proof}
    Suppose we use a $\m N(0,\sigma^2_\beta)$ prior for $\beta$ for $\sigma_\beta= \sqrt{\log n}$. Then $|\beta^*| \ll |\log(\epsilon_n^2)| \lesssim |\log(\log(n)/n^2))| \lesssim \log n= \sigma_{\beta}^2$.  Then, the prior concentration shows that 
 	\begin{equation*}
	\begin{aligned}
	 & \Pi\left(|\beta-\beta^*|\leq \epsilon  \right) \\
	&\gtrsim \frac{1}{(\sqrt{2\pi}\sigma_\beta)}\exp(- \frac{\beta^{*2} }{2\sigma_\beta^{2}}) (2\epsilon)   \\
	& \gtrsim \exp \left[-K \left\{\frac{(\beta^*)^2}{2 \sigma_\beta^{2}}+ \log( \frac{1}{\epsilon})+\log (\sigma_\beta) \right\}\right].
	\end{aligned}
	\end{equation*}
Then $\beta^{*2}/\sigma_\beta \lesssim  |\beta^{*}| \lesssim \log(1/\epsilon_n)$ and we have $\log(1/\epsilon_{n}) \lesssim n^2\epsilon_n^2$ for any $s_b,s_x$, we can see the logarithm of the right-hand side of the above inequality is lower bounded by $-K_0n^2\epsilon_{n}^2$ for some constant $K_0>0$.

\end{proof}
	\begin{lemma}[Upper bound for binary KL divergence]\label{lem:binary_KL}
		Let $p_a=1/(1+\exp(-a))$ and $p_b=1/(1+\exp (-b))$. Define $P_a$ and $P_b$ as the Bernoulli measures with probability $p_a$ and $p_b$. Then we have
		\begin{equation*}
		D_{KL}(P_{a} \,||\, P_{b}) + D_{KL}(P_{b} \,||\, P_{a}) \leq (p_a \vee p_b)(a-b)^2.
		\end{equation*}
	\end{lemma}
	
	\begin{proof}
		\begin{equation*}
		\begin{aligned}
		D_{KL}(P_{a} \,||\, P_{b}) + D_{KL}(P_{b} \,||\, P_{a}) &=(p_a-p_b) \log \frac{p_a}{p_b}+(p_b-p_a) \log \frac{1-p_a}{1-p_b}\\
		=(p_a-p_b) \log\left(\frac{p_a}{1-p_a} \frac{1-p_b}{p_b}\right) 
		&= \left\{\frac{1}{1+\exp(-a)}-\frac{1}{1+\exp(-b)}\right\} \log(e^{a} e^{-b}) \\
		& = (a-b)\left\{\frac{1}{1+\exp(-a)}-\frac{1}{1+\exp(-b)}\right\}.
		\end{aligned}
		\end{equation*}
		Without loss of generality, we can assume $a>b$, then by $\exp(x)\geq 1+x$, we have
		\begin{align*}
		\frac{1}{1+\exp(-a)}-\frac{1}{1+\exp(-b)} = \frac{e^{-b}-e^{-a}}{(1+\exp(-a))(1+\exp(-b))} \\
		\leq  \frac{1-e^{b-a}}{(1+e^{-a})(1+e^b)} \leq p_a(1-e^{b-a}) \leq p_a(a-b).
		\end{align*}
	\end{proof}

	\begin{lemma}[Upper bound of second order KL moment]\label{lem:second_KL}
		Let $p_a=1/(1+\exp(-a))$ and $p_b=1/(1+\exp (-b))$. Define $P_a$ and $P_b$ as the Bernoulli measures with probability $p_a$ and $p_b$. Then we have
		\begin{equation*}
		\int P_a \log^2 \left( \frac{P_a}{P_b}\right) d\mu \leq  \left[\frac{p_a}{(p_a \wedge  p_b)^2}+\frac{1-p_a}{(1-p_a\vee p_b)^2}\right] (p_a \vee p_b)^2(a-b)^2.
		\end{equation*}
	\end{lemma}
 
\begin{proof}
 Note that 
		\begin{equation*}
		\begin{aligned}
\int P_a \log^2 \left( \frac{P_a}{P_b}\right) d\mu = p_a\log^2 \left( \frac{p_a}{p_b}\right) +(1-p_a) \log^2\left( \frac{1-p_a}{1-p_b}\right).
		\end{aligned}
		\end{equation*}
  We have
\begin{align*}
    \log^2 \left( \frac{p_a}{p_b}\right)  = \log^2\left( \frac{p_a \vee p_b}{p_a \wedge p_b}-1+1\right) \leq \left(\frac{p_a \vee p_b - p_a \wedge p_b}{p_a \wedge p_b}\right)^2 = \left(\frac{p_a - p_b}{p_a \wedge p_b}\right)^2.
\end{align*}
  Similarly,
\begin{align*}
    \log^2 \left( \frac{1-p_a}{1-p_b}\right)  = \log^2\left( \frac{(1-p_a) \vee (1-p_b)}{(1-p_a) \wedge (1-p_b)}-1+1\right) \leq \left(\frac{p_a - p_b}{1-p_a \vee p_b}\right)^2.
\end{align*}
  
	For the $(p_a-p_b)^2$ term,  by $\exp(x)\geq 1+x$, we have
		\begin{align*}
		\frac{1}{1+\exp(-a \vee b)}-\frac{1}{1+\exp(-a \wedge b)} = \frac{e^{-a \wedge b}-e^{-a \vee b}}{(1+\exp(-a \vee b))(1+\exp(-a \wedge b))} \\
		\leq  \frac{1-e^{a \vee b-a \wedge b}}{(1+e^{a \wedge b})(1+e^{-a \vee b})} \leq (p_a \vee p_b)(1-e^{a \vee b-a \wedge b}) \leq (p_a \vee p_b) (a \wedge b-a \vee b).
		\end{align*}
	\end{proof}

	\begin{lemma}[Lower bound of the $1/2$ divergence] \label{lem:lower of divergence}Let $p_a=1/(1+\exp(-a))$ and $p_b=1/(1+\exp (-b))$ with $|a|,|b| \gtrsim 1$ and $a,b<0$. Define $P_a$ and $P_b$ as the Bernoulli measures with probability $p_a$ and $p_b$.
 Then we have
		\begin{equation*}
		D_{\frac{1}{2}}(P_a,P_b) \gtrsim  \exp \left\{a \wedge b \right\}(b-a)^2.
		\end{equation*}

  	\end{lemma}
  \begin{proof}
   		\begin{align*}
		D_{\frac{1}{2}}(P_a,P_b)=-2\log(1-h^2(p_a,p_b)) \geq 2h^2(p_a,p_b) =  \left[(\sqrt{p_a}-\sqrt{p_b})^2+(\sqrt{1-p_a}-\sqrt{1-p_b})^2\right].
		\end{align*}
      For the term $(\sqrt{p_a}-\sqrt{p_b})^2$, by the mean value theorem of function $\sqrt{p_x}$ with respect to $x$, we have
  \begin{equation*}
     (\sqrt{p_a}-\sqrt{p_b})^2 \geq  \left( \frac{\sqrt{\exp(x)} \sqrt{1+\exp(x)}}{2 (1+\exp(x))^2} \right)^2 (a-b)^2 \stackrel{(i)}{\gtrsim} e^{a \wedge b }(a-b)^2,
  \end{equation*}
  where $(i)$ is because $\exp(x)$ is the order of $\exp \{a \wedge b \}$ for $a\wedge b<x<a \vee b$. For the term $(\sqrt{1-p_a}-\sqrt{1-p_b})^2$, note that $\sqrt{1-p_a}+\sqrt{1-p_b}$ is still bound away from $0$, we have 
  \begin{equation*}
      (\sqrt{1-p_a}-\sqrt{1-p_b})^2 \gtrsim (p_a-p_b)^2 = \left\{\frac{\exp(x)}{(1+\exp(x))^2} \right\}^2(a-b)^2 \gtrsim e^{(2a) \wedge (2b) } (a-b)^2,
  \end{equation*}
  for $a<x<b$. Finally, $\exp(a \wedge b )(a-b)^2$ dominates when the sum of the two lower bounds is taken into account.
  \end{proof}
% \begin{lemma}[Jensen gap for the Sigmoid function]
% Consider $n$ values $x_1, x_2, \ldots, x_n$ such that we have $\sum_{i=1}^n x_i/n = \bar x$, $\max_i\|x_i-\bar x\|\lesssim 1$ and $\bar x \rightarrow -\infty$. Then we have the Jensen gap
% \begin{equation*}
%     \frac{1}{n} \sum_{i=1}^n S(x_i) - S(\bar x) \lesssim S(\bar x).
% \end{equation*}
% \end{lemma}
% \begin{proof}
% By Taylor's expansion of the Sigmoid function,
% \begin{equation*}
%     S(x_i) = S(\bar x) + S'(\bar x) (x_i - \bar x) +\frac{S^{''}(\tilde{x}_i)}{2} (x_i - \bar x)^2,
% \end{equation*}
% where $\tilde{x}_i$ is a value between $x_i$ and $\bar x$. Then,
% \begin{equation*}
%      \frac{1}{n} \sum_{i=1}^n S(x_i) - S(\bar x) =\sum_{i=1}^n \frac{S^{''}(\tilde{x}_i)}{2n}(x_i - \bar x)^2 \lesssim S^{''}(\bar x),
% \end{equation*}
% where the last inequality is because $S^{''}(\tilde{x}_i) \lesssim S''(\bar x)$ when $\max_i\|x_i-\bar x\|\lesssim 1$ and then use the form of the second order derivative of the Sigmoid function.
% \end{proof}

  	\begin{lemma}[Perturbation bound for Procrustes, Lemma 5.4 in \cite{tu2016low}] \label{lem:perturbation}
	 For any $\+{U}_1, \+{U}_2 \in \mathbb{R}^{n \times d}$, we have
$$
\min _{\+{O} \in \mathcal{O}_d}\left\|\+{U}_1-\+{U}_2 \+{O}\right\|_F^2 \leq \frac{1}{2(\sqrt{2}-1) \sigma_d^2\left(\+{U}_2\right)}\left\|\+{U}_1 \+{U}_1^{'}-\+{U}_2 \+{U}_2^{'}\right\|_F^2,
$$
where $\sigma_d\left(\+{U}_2\right)$ is the $d$-th largest singular value of $\+{U}_2$.
	\end{lemma}

	\begin{lemma}[$K$-means error bound, adapted from Lemma 5.3 in  \cite{lei2015consistency}] \label{lem:k_means}
		For any two matrices $\hat{\+U},\+U\in \mathbb{R}^{n \times d}$ such that $\+U=\+\Theta \+X$ with $\+\Theta \in \mathbb{M}_{n,K}$, $\+X \in \mathbb{R}^{K \times d}$, let $(\hat{\+\Theta},\hat{\+X})$ be the solution to the $K$-means problem and $\bar{\+U}=\hat{\+\Theta}\hat{\+X}$. Then for any $\delta_k \leq \min_{l \ne k} \|\+u_{l}-\+u_{k} \|_2$, define $S_k= \{i \in G_k(\+\Theta): \|\bar{\+u}_i-\+u_{i} \|_2 \geq \delta_k/2\}$, then
		\begin{equation*}
		\sum_{k=1}^{K} |S_k|\delta_k^2 \leq {16 \|\hat{\+U}-\+U\|_F^2}.
		\end{equation*}
	\end{lemma}
	Moreover, if
	\begin{equation*}
	16 \|\hat{\+U}-\+U\|_F^2/\delta_k^2 <n_k \quad \text{for all } k,
	\end{equation*}
	then there exists a $K \times K$ permutation matrix $J$ such that $\hat{\+\Theta}_{G^*}=\+\Theta_{G^*} J$, where $G = \cap_{k=1}^K(G_k-S_k)$.

% {\putbib}
% 	\end{bibunit}

\bibliographystyle{apalike}
\bibliography{main}

\begin{thebibliography}{}

\bibitem[Akimushkin et~al., 2017]{akimushkin2017text}
Akimushkin, C., Amancio, D.~R., and Oliveira~Jr, O.~N. (2017).
\newblock Text authorship identified using the dynamics of word co-occurrence
  networks.
\newblock {\em PloS one}, 12(1):e0170527.

\bibitem[Bhattacharya et~al., 2019]{bhattacharya2019bayesian}
Bhattacharya, A., Pati, D., and Yang, Y. (2019).
\newblock Bayesian fractional posteriors.
\newblock {\em The Annals of Statistics}, 47(1):39--66.

\bibitem[Bickel et~al., 2009]{bickel2009simultaneous}
Bickel, P.~J., Ritov, Y., and Tsybakov, A.~B. (2009).
\newblock Simultaneous analysis of lasso and dantzig selector.
\newblock {\em The Annals of Statistics}, pages 1705--1732.

\bibitem[Binkiewicz et~al., 2017]{binkiewicz2017covariate}
Binkiewicz, N., Vogelstein, J.~T., and Rohe, K. (2017).
\newblock Covariate-assisted spectral clustering.
\newblock {\em Biometrika}, 104(2):361--377.

\bibitem[Bishop and Nasrabadi, 2006]{bishop2006pattern}
Bishop, C.~M. and Nasrabadi, N.~M. (2006).
\newblock {\em Pattern recognition and machine learning}, volume~4.
\newblock Springer.

\bibitem[Blei et~al., 2017]{blei2017variational}
Blei, D.~M., Kucukelbir, A., and McAuliffe, J.~D. (2017).
\newblock Variational inference: A review for statisticians.
\newblock {\em Journal of the American Statistical Association},
  112(518):859--877.

\bibitem[Carvalho et~al., 2010]{carvalho2010horseshoe}
Carvalho, C.~M., Polson, N.~G., and Scott, J.~G. (2010).
\newblock The horseshoe estimator for sparse signals.
\newblock {\em Biometrika}, 97(2):465--480.

\bibitem[Csardi, 2015]{Csardi2015igraphdata}
Csardi, G. (2015).
\newblock {\em igraphdata: A Collection of Network Data Sets for the 'igraph'
  Package}.
\newblock R package version 1.0.1.

\bibitem[Eddelbuettel and Fran\c{c}ois, 2011]{Eddelbuettel2011Rcpp}
Eddelbuettel, D. and Fran\c{c}ois, R. (2011).
\newblock {Rcpp}: Seamless {R} and {C++} integration.
\newblock {\em Journal of Statistical Software}, 40(8):1--18.

\bibitem[Fortunato and Hric, 2016]{fortunato2016community}
Fortunato, S. and Hric, D. (2016).
\newblock Community detection in networks: A user guide.
\newblock {\em Physics reports}, 659:1--44.

\bibitem[Goldenberg et~al., 2010]{goldenberg2010survey}
Goldenberg, A., Zheng, A.~X., Fienberg, S.~E., and Airoldi, E.~M. (2010).
\newblock A survey of statistical network models.

\bibitem[Hager and Talbert, 2000]{hager2000look}
Hager, G.~L. and Talbert, J.~C. (2000).
\newblock Look for the party label: Party influences on voting in the us house.
\newblock {\em Legislative Studies Quarterly}, pages 75--99.

\bibitem[Handcock et~al., 2007]{handcock2007model}
Handcock, M.~S., Raftery, A.~E., and Tantrum, J.~M. (2007).
\newblock Model-based clustering for social networks.
\newblock {\em Journal of the Royal Statistical Society: Series A (Statistics
  in Society)}, 170(2):301--354.

\bibitem[Hoff, 2008]{NIPS2007_766ebcd5}
Hoff, P. (2008).
\newblock Modeling homophily and stochastic equivalence in symmetric relational
  data.
\newblock In Platt, J., Koller, D., Singer, Y., and Roweis, S., editors, {\em
  Advances in Neural Information Processing Systems}, volume~20, pages
  657--664. Curran Associates, Inc.

\bibitem[Hoff et~al., 2002]{hoff2002latent}
Hoff, P.~D., Raftery, A.~E., and Handcock, M.~S. (2002).
\newblock Latent space approaches to social network analysis.
\newblock {\em Journal of the American Statistical Association},
  97(460):1090--1098.

\bibitem[Hu and Wang, 2024]{hu2024network}
Hu, Y. and Wang, W. (2024).
\newblock Network-adjusted covariates for community detection.
\newblock {\em Biometrika}, page asae011.

\bibitem[Jaakkola and Jordan, 2000]{jaakkola2000bayesian}
Jaakkola, T.~S. and Jordan, M.~I. (2000).
\newblock {Bayesian} parameter estimation via variational methods.
\newblock {\em Statistics and Computing}, 10(1):25--37.

\bibitem[Jeong and Ghosal, 2021]{jeong2021posterior}
Jeong, S. and Ghosal, S. (2021).
\newblock Posterior contraction in sparse generalized linear models.
\newblock {\em Biometrika}, 108(2):367--379.

\bibitem[Jin et~al., 2019]{jin2019short}
Jin, C., Netrapalli, P., Ge, R., Kakade, S.~M., and Jordan, M.~I. (2019).
\newblock A short note on concentration inequalities for random vectors with
  subgaussian norm.
\newblock {\em arXiv preprint arXiv:1902.03736}.

\bibitem[Krivitsky et~al., 2009]{krivitsky2009representing}
Krivitsky, P.~N., Handcock, M.~S., Raftery, A.~E., and Hoff, P.~D. (2009).
\newblock Representing degree distributions, clustering, and homophily in
  social networks with latent cluster random effects models.
\newblock {\em Social networks}, 31(3):204--213.

\bibitem[Lei and Rinaldo, 2015]{lei2015consistency}
Lei, J. and Rinaldo, A. (2015).
\newblock Consistency of spectral clustering in stochastic block models.
\newblock {\em The Annals of Statistics}, 43(1):215--237.

\bibitem[Li et~al., 2023]{Li2023randnet}
Li, T., Levina, E., Zhu, J., and {Can M. Le} (2023).
\newblock {\em randnet: Random Network Model Estimation, Selection and
  Parameter Tuning}.
\newblock R package version 0.7.

\bibitem[Liu and Chen, 2022]{liu2022variational}
Liu, Y. and Chen, Y. (2022).
\newblock Variational inference for latent space models for dynamic networks.
\newblock {\em Statistica Sinica}, 32(4):2147--2170.

\bibitem[Loshchilov and Hutter, 2019]{loshchilov2019decoupled}
Loshchilov, I. and Hutter, F. (2019).
\newblock Decoupled weight decay regularization.
\newblock In {\em International Conference on Learning Representations}.

\bibitem[Loyal, 2024]{loyal2024fast}
Loyal, J.~D. (2024).
\newblock Fast variational inference of latent space models for dynamic
  networks using {Bayesian} p-splines.
\newblock {\em arXiv preprint arXiv:2401.09715}.

\bibitem[Loyal and Chen, 2023]{loyal2023eigenmodel}
Loyal, J.~D. and Chen, Y. (2023).
\newblock An eigenmodel for dynamic multilayer networks.
\newblock {\em The Journal of Machine Learning Research}, 24(128):1--69.

\bibitem[Ma et~al., 2020]{ma2020universal}
Ma, Z., Ma, Z., and Yuan, H. (2020).
\newblock Universal latent space model fitting for large networks with edge
  covariates.
\newblock {\em The Journal of Machine Learning Research}, 21(4):1--67.

\bibitem[Martin and Tang, 2020]{martin2020empirical}
Martin, R. and Tang, Y. (2020).
\newblock Empirical priors for prediction in sparse high-dimensional linear
  regression.
\newblock {\em The Journal of Machine Learning Research}, 21(1):5709--5738.

\bibitem[Mart{\'\i}nez et~al., 2016]{martinez2016survey}
Mart{\'\i}nez, V., Berzal, F., and Cubero, J.-C. (2016).
\newblock A survey of link prediction in complex networks.
\newblock {\em ACM computing surveys (CSUR)}, 49(4):1--33.

\bibitem[McCallum et~al., 2000]{mccallum2000automating}
McCallum, A.~K., Nigam, K., Rennie, J., and Seymore, K. (2000).
\newblock Automating the construction of internet portals with machine
  learning.
\newblock {\em Information Retrieval}, 3:127--163.

\bibitem[Mitchell and Beauchamp, 1988]{mitchell1988bayesian}
Mitchell, T.~J. and Beauchamp, J.~J. (1988).
\newblock Bayesian variable selection in linear regression.
\newblock {\em Journal of the american statistical association},
  83(404):1023--1032.

\bibitem[Namata et~al., 2012]{namata2012query}
Namata, G., London, B., Getoor, L., Huang, B., and Edu, U. (2012).
\newblock Query-driven active surveying for collective classification.
\newblock In {\em 10th international workshop on mining and learning with
  graphs}, volume~8, page~1.

\bibitem[Neville et~al., 2014]{neville2014mean}
Neville, S.~E., Ormerod, J.~T., and Wand, M. (2014).
\newblock Mean field variational {Bayes} for continuous sparse signal
  shrinkage: pitfalls and remedies.
\newblock {\em Electronic Journal of Statistics}, 8(1):1113--1151.

\bibitem[Newman, 2018]{newman2018networks}
Newman, M. (2018).
\newblock {\em Networks}.
\newblock Oxford university press.

\bibitem[Newman and Clauset, 2016]{newman2016structure}
Newman, M.~E. and Clauset, A. (2016).
\newblock Structure and inference in annotated networks.
\newblock {\em Nature communications}, 7(1):11863.

\bibitem[Niu et~al., 2023]{niu2023covariate}
Niu, Y., Ni, Y., Pati, D., and Mallick, B.~K. (2023).
\newblock Covariate-assisted {Bayesian graph} learning for heterogeneous data.
\newblock {\em Journal of the American Statistical Association}, pages 1--15.

\bibitem[Rockova and George, 2018]{rovckova2018spike}
Rockova, V. and George, E.~I. (2018).
\newblock The spike{-}and{-}slab lasso.
\newblock {\em Journal of the American Statistical Association},
  113(521):431--444.

\bibitem[Rowe et~al., 2007]{rowe2007automated}
Rowe, R., Creamer, G., Hershkop, S., and Stolfo, S.~J. (2007).
\newblock Automated social hierarchy detection through email network analysis.
\newblock In {\em Proceedings of the 9th WebKDD and 1st SNA-KDD 2007 workshop
  on Web mining and social network analysis}, pages 109--117.

\bibitem[Rozemberczki and Sarkar, 2020]{rozemberczki2020characteristic}
Rozemberczki, B. and Sarkar, R. (2020).
\newblock {Characteristic Functions on Graphs: Birds of a Feather, from
  Statistical Descriptors to Parametric Models}.
\newblock In {\em Proceedings of the 29th ACM International Conference on
  Information and Knowledge Management (CIKM '20)}, page 1325–1334. ACM.

\bibitem[Sewell and Chen, 2017]{sewell2017latent}
Sewell, D.~K. and Chen, Y. (2017).
\newblock Latent space approaches to community detection in dynamic networks.
\newblock {\em Bayesian analysis}, 12(2):351--377.

\bibitem[Snijders, 2011]{snijders2011statistical}
Snijders, T.~A. (2011).
\newblock Statistical models for social networks.
\newblock {\em Annual review of sociology}, 37:131--153.

\bibitem[Stopczynski and Lehmann, 2018]{stopczynski2018physical}
Stopczynski, A. and Lehmann, S. (2018).
\newblock How physical proximity shapes complex social networks.
\newblock {\em Scientific reports}, 8(1):1--10.

\bibitem[Tu et~al., 2016]{tu2016low}
Tu, S., Boczar, R., Simchowitz, M., Soltanolkotabi, M., and Recht, B. (2016).
\newblock Low-rank solutions of linear matrix equations via procrustes flow.
\newblock In {\em International Conference on Machine Learning}, pages
  964--973. PMLR.

\bibitem[Vershynin, 2010]{vershynin2010introduction}
Vershynin, R. (2010).
\newblock Introduction to the non-asymptotic analysis of random matrices.
\newblock {\em arXiv preprint arXiv:1011.3027}.

\bibitem[Von~Mering et~al., 2002]{von2002comparative}
Von~Mering, C., Krause, R., Snel, B., Cornell, M., Oliver, S.~G., Fields, S.,
  and Bork, P. (2002).
\newblock Comparative assessment of large-scale data sets of protein--protein
  interactions.
\newblock {\em Nature}, 417(6887):399--403.

\bibitem[Walker and Hjort, 2001]{walker2001bayesian}
Walker, S. and Hjort, N.~L. (2001).
\newblock On {Bayesian} consistency.
\newblock {\em Journal of the Royal Statistical Society: Series B (Statistical
  Methodology)}, 63(4):811--821.

\bibitem[Yan and Sarkar, 2021]{yan2021covariate}
Yan, B. and Sarkar, P. (2021).
\newblock Covariate regularized community detection in sparse graphs.
\newblock {\em Journal of the American Statistical Association},
  116(534):734--745.

\bibitem[Yan et~al., 2019]{yan2019statistical}
Yan, T., Jiang, B., Fienberg, S.~E., and Leng, C. (2019).
\newblock Statistical inference in a directed network model with covariates.
\newblock {\em Journal of the American Statistical Association},
  114(526):857--868.

\bibitem[Zhang et~al., 2022a]{zhang2022directed}
Zhang, J., He, X., and Wang, J. (2022a).
\newblock Directed community detection with network embedding.
\newblock {\em Journal of the American Statistical Association},
  117(540):1809--1819.

\bibitem[Zhang et~al., 2022b]{zhang2022joint}
Zhang, X., Xu, G., and Zhu, J. (2022b).
\newblock Joint latent space models for network data with high-dimensional node
  variables.
\newblock {\em Biometrika}, 109(3):707--720.

\bibitem[Zhang et~al., 2020]{zhang2020flexible}
Zhang, X., Xue, S., and Zhu, J. (2020).
\newblock A flexible latent space model for multilayer networks.
\newblock In {\em International Conference on Machine Learning}, pages
  11288--11297. PMLR.

\bibitem[Zhang et~al., 2016]{zhang2016community}
Zhang, Y., Levina, E., and Zhu, J. (2016).
\newblock Community detection in networks with node features.
\newblock {\em Electronic Journal of Statistics}, 10(2):3153--3178.

\bibitem[Zhao et~al., 2022a]{zhao2022factorized}
Zhao, P., Bhattacharya, A., Pati, D., and Mallick, B.~K. (2022a).
\newblock Factorized fusion shrinkage for dynamic relational data.
\newblock {\em arXiv preprint arXiv:2210.00091}.

\bibitem[Zhao et~al., 2022b]{zhao2022structured}
Zhao, P., Bhattacharya, A., Pati, D., and Mallick, B.~K. (2022b).
\newblock Structured optimal variational inference for dynamic latent space
  models.
\newblock {\em arXiv preprint arXiv:2209.15117}.

\end{thebibliography}

\end{document}